\documentclass[8pt,onefignum,onetabnum]{extarticle}

\usepackage{amsfonts}
\usepackage{amsmath}
\usepackage{amssymb}
\usepackage{amsthm}
\usepackage{graphicx}
\usepackage{epstopdf}
\usepackage{subcaption}
\usepackage{verbatim}
\usepackage{esvect}
\usepackage{xcolor}
\usepackage{caption}
\usepackage{float}
\usepackage{multirow}
\usepackage{multicol}
\usepackage[utf8]{inputenc}
\usepackage[linesnumbered,ruled,vlined]{algorithm2e}
\usepackage{hyperref}
\usepackage{cleveref}

\SetKwComment{Comment}{\# }{}
\SetNlSty{}{}{}
\let\oldnl\nl
\newcommand\nonl{%
  \renewcommand{\nl}{\let\nl\oldnl}}
\SetCommentSty{mycommfont}
\ifpdf
  \DeclareGraphicsExtensions{.eps,.pdf,.png,.jpg}
\else
  \DeclareGraphicsExtensions{.eps}
\fi
\SetKwInput{KwInput}{Input}                
\SetKwInput{KwOutput}{Output}              

\makeatletter
\def\hlinewd#1{%
\noalign{\ifnum0=`}\fi\hrule \@height #1 \futurelet
\reserved@a\@xhline}
\makeatother

\newcommand{\email}[1]{\texttt{#1}}
\newtheorem{theorem}{Theorem}[section]

\title{Probing for the Trace Estimation of a Permuted Matrix Inverse Corresponding to a Lattice Displacement\thanks{This work is supported by SURA under grant number C2019-FEMT-002-05, by the Exascale Computing Project (17-SC-20-SC), a collaborative effort of the U.S. Department of Energy Office of Science and the National Nuclear Security Administration, and by Deakin University. 
The authors acknowledge William \& Mary Research Computing for providing computational resources that have contributed to the results reported within this paper (\url{https://www.wm.edu/it/rc}). }}

\author{Heather Switzer\thanks{Department of Computer Science, College of William \& Mary, Williamsburg, VA (\email{hmswitzer@wm.edu}).}
\and Andreas Stathopoulos\thanks{Department of Computer Science, College of William \& Mary, Williamsburg, VA (\email{andreas@cs.wm.edu}).}
\and Eloy Romero\thanks{Jefferson Laboratory, Newport News, VA (\email{eromero@jlab.org}).}
\and Jesse Laeuchli\thanks{School of Information Technology, Deakin University, Geelong, Victoria 3220, Australia (\email{j.laeuchli@deakin.edu.au}).}
\and Kostas Orginos \thanks{Jefferson Laboratory, Newport News, VA (\email{knorgi@wm.edu}).}
}

\usepackage{amsopn}

\makeatletter
\newcommand*{\addFileDependency}[1]{
  \typeout{(#1)}
  \@addtofilelist{#1}
  \IfFileExists{#1}{}{\typeout{No file #1.}}
}
\makeatother

\begin{document}

\maketitle

\begin{abstract}
Probing \cite{Probing} is a general technique that is used to reduce the variance of the Hutchinson stochastic estimator for the trace of the inverse of a large, sparse matrix $A$ \cite{Hutchinson}. The variance of the estimator is the sum of the squares of the off-diagonal elements of $A^{-1}$. Therefore, this technique computes probing vectors that when used in the estimator they annihilate the largest off-diagonal elements. 
For matrices that display decay of the magnitude of $|A^{-1}_{ij}|$ with the graph distance between nodes $i$ and $j$, this is achieved through graph coloring of increasing powers $A^k$ \cite{Poly}. Equivalently, when a matrix stems from a lattice discretization, it is computationally beneficial to find a distance-$k$ coloring of the lattice. In \cite{HP} a hierarchical coloring was proposed so that $k$ can be increased at runtime as needed without discarding previous work.

In this work, we study probing for the more general problem of computing the trace of a permutation of $A^{-1}$, say $PA^{-1}$. 
The motivation comes from Lattice QCD where we need to construct ``disconnected diagrams'' to extract flavor-separated Generalized Parton functions. In Lattice QCD, where the matrix has a 4D toroidal lattice structure, these non-local operators correspond to a $PA^{-1}$ where $P$ is the permutation relating to some displacement $\vec{p}$ in one or more dimensions. We focus on a single dimension displacement ($p$) but our methods are general.
We show that probing on $A^k$ or $(PA)^k$ do not annihilate the largest magnitude elements. To resolve this issue, our displacement-based probing works on $PA^k$ using a new coloring scheme that works directly on appropriately displaced neighborhoods on the lattice. We prove lower bounds on the number of colors needed, and study the effect of this scheme on variance reduction, both theoretically and experimentally on a real-world Lattice QCD calculation. We achieve orders of magnitude speedup over the unprobed or the naively probed methods.

\end{abstract}



\section{Introduction}
\label{sec:intro}

The approximation of the trace of a matrix function, $f(A)$, of a large sparse matrix $A$ is a computationally challenging problem. Commonly used functions are the $A^{-1}$ and $\log A$ (which is used to find the matrix determinant). In this paper we focus on $f(A)=A^{-1}$ which has many applications in statistics \cite{Hutchinson}, quantum Monte Carlo \cite{qMC}, and data mining \cite{DataMining}. 
Our motivating application comes from lattice quantum chromodynamics (LQCD). In LQCD, the trace of the inverse of an operator discretized on a symmetric, four-dimensional, toroidal lattice representing space-time is often used to analyze the interactions, properties, and structure of hadrons on a subatomic scale \cite{WhatIsLQCD}. The trace computations are part of larger scale Monte Carlo simulations and therefore do not require high accuracy but must induce no statistical bias. 

Effective methods for computing $\text{\textbf{Tr}}(A^{-1})$ exist for smaller matrices where sparse factorizations are possible or in cases where selective elements of the inverse can be found \cite{LU1, LU2, Lin-Chao_traceInv}. However, as the size and density of $A$ increases they become computationally infeasible and stochastic estimation is the only alternative. A widely used method for this is the Hutchinson's trace estimator \cite{Hutchinson} which takes the form
\begin{equation}
    \text{\textbf{Tr}}(A^{-1}) \approx \frac{1}{s}\sum_{i=1}^s z_i^T A^{-1} z_i,
    \label{eq:Hutchinson}
\end{equation}
where $z_i$ are $s$ i.i.d. random noise vectors (RNV). 
The computational complexity therefore is dominated by the solution of the linear systems with some iterative method to approximate the Guassian quadrature $z_i^TA^{-1}z_i$ at each step.
The RNVs are chosen to have a Rademacher distribution, where each element is equal to $\pm 1$ with probability 0.5. It is known that for this choice the estimator has variance 
\begin{equation}
    \text{Var}(z^TA^{-1}z) = 2(\| A^{-1}\|^2_F - \sum_{i=1}^N(A_{i,i})^2),
    \label{eq:HutchinsonVar}
\end{equation}
which is minimum over all random distributions for $z_i$ when $A$ is real \cite{Struct1}. The variance is the same for complex matrices, which is the case in LQCD, when $\mathbb{Z}_4$ Rademacher vectors are used, i.e., vectors with $\pm 1, \pm i$ values with probability 0.25.
The variance formula shows that large off-diagonal elements contribute significant errors to the estimator and cause slow convergence. Many techniques have been introduced and studied to reduce the variance of the estimator by choosing vectors that better take advantage of the structure of the matrix \cite{Struct1, DataMining, Struct2, Probing, HadamardSampling}.

One such technique is classical probing (CP). Probing is a general technique that uses graph coloring of the graph of an adjacency matrix $A$ to construct structurally orthogonal probing vectors to extract specific non-zero entries of the matrix. For example, multiplying a diagonal matrix with a vector of ones recovers its diagonal. Similarly, when the adjacency matrix of a graph is $m$-colorable, we can also recover the diagonal by multiplying the matrix with $m$ vectors, each vector having ones in rows with the same color and zero elsewhere. In numerical optimization probing is applied on the graph of $A^2$ in order to compute the Hessian 
\cite{Pothen_Coloring}. 
For trace estimation, CP constructs probing vectors from a coloring of the graph of $A^k$ or equivalently the distance-$k$ coloring of the graph of $A$, where $k \in \mathbb{Z}_+$ \cite{Probing}. The idea is that for many sparse matrices the elements of $A^{-1}_{ij}$ display a Green's function decay in magnitude with the distance between nodes $i$ and $j$. Although $A^{-1}$ is not sparse, using these probing vectors in the estimator removes from the variance (\ref{eq:HutchinsonVar}) all elements (edges) of neighbors with distance $\leq k$. A drawback of CP is that if a coloring for a certain distance $k$ does not produce the required variance reduction, a higher distance coloring cannot reuse the quadratures computed with the previous probing vectors.

Hierarchical probing (HP) was introduced to address the reuse issue \cite{HP, EHP}. HP assigns colors to nodes in a hierarchical way so that two nodes that receive the same color for some distance $k$ will never share the same color in higher distances. The technique also provided a computationally inexpensive way to produce a distance-$k$ coloring for large $k$ when the matrix graph is a regular, toroidal lattice. This toroidal structure appears in LQCD matrices which is also the focus of the current paper. 

Deflation has also been used as a variance reduction technique \cite{DeflationMG, Deflation, Guerrero:2010}. While probing techniques capture large elements from relatively small lattice distances, the low rank approximation of $A^{-1}$ using the lowest magnitude singular triplets of $A$ typically captures a large part of the magnitude of $A^{-1}$ at long distances.
Thus, the two approaches are complementary and, when used in tandem, can significantly accelerate the Monte Carlo estimator.

In this paper we extend probing for computing the trace of a permutation of $A^{-1}$. The motivation comes from LQCD computations of the flavor-separated Generalized Parton functions (GPDs) where the so-called ``disconnected diagrams'' need to be calculated \cite{DisDiag,Alexandrou:2020uyt}. This translates to the need to find the sum of certain off-diagonal elements of $A^{-1}$ that correspond to a displacement along the $z$ dimension of the four-dimensional (space-time) LQCD lattice.
This is a non-symmetric permutation of the rows of $A^{-1}$, where the index of a node $x$ no longer refers to $[x_1, x_2, x_3, x_4]$, but instead $[x_1, x_2, x_3+p, x_4]$. The associated trace problem is more challenging because the variance for $PA^{-1}$ now includes the main diagonal $A^{-1}$ which is of much larger magnitude than the one of $PA^{-1}$.

We propose an extension of CP that modifies a greedy coloring algorithm to consider not the node's original neighborhood but the neighborhood of its displacement. The idea applies to any permutation matrix and can be performed in a hierarchical way if desired. For toroidal lattices with displacement applied in one dimension we prove lower bounds on the number of colors and study the effect of the algorithm on variance reduction both theoretically and with LQCD experiments. The method results in orders of magnitude variance reduction over conventional probing methods.

The rest of the paper is organized as follows: \Cref{sec:background}
introduces notation and discusses previous variance reduction techniques. \Cref{sec:ProbwDis} introduces the coloring algorithm with displacements, and studies its properties theoretically. Experimental result are shown in \Cref{sec:Experiments}. Conclusions and some open questions are given in \Cref{sec:Conclusion}.

\section{Background}
\label{sec:background}

In this paper we seek the trace of $PA^{-1}$, where $P$ is a permutation matrix, and $A$ is a non-singular matrix of dimension $N$.  $A$ can be complex valued as in the case of LQCD but for convenience and without loss of generality our presentation involves real valued matrices. Although our main idea applies to any $P$ and $A$, the algorithm and the analysis is relevant to matrices stemming from a regular lattice discretization.
Letting $\mathbb{Z}_n$ be the multiplicative group of integers modulo $n$, then a $d$-dimensional toroidal lattice is described as
\begin{equation}
    \mathbb{Z}^d_D = \mathbb{Z}_{D_1} \times ... \times \mathbb{Z}_{D_d},
\end{equation}
where $D_i$ is the size of dimension $i$. Two lattice nodes $x$ and $y$ are connected by an edge if their coordinate vectors $[x_1, ..., x_d]$ and $[y_1, ..., y_d]$, satisfy $\| x - y\|_1 = 1$ (in a modulo sense). In LQCD, the lattice represents the 4 dimensional space-time. 

Variance reduction techniques for the Hutchinson trace estimator focus around two approaches; one derives an approximation to $A^{-1}$ such as from deflation or preconditioning which we brief\/ly address in \Cref{sec:deflation}; the other replaces the Rademacher vectors with ones that better take advantage of the structure of the matrix. Orthogonal columns of the Hadamard or Fourier matrix have been proposed \cite{DataMining} which can systematically annihilate specific diagonals of the matrix and thus reduce the variance in (\ref{eq:HutchinsonVar}). 
The variance reduction is monotonic with the number of columns used but this method works no better than using solely RNVs as the patterns of diagonals removed are not typically the heaviest variance-contributing diagonals of $A^{-1}$. The following methods attempt to capture these heaviest elements directly.

\subsection{Classical Probing}

The inverse of an $N \times N$ non-singular matrix $A$ where $\| A \| < 1$ can be represented by the Neumann series $A^{-1} = \sum_{k=0}^{\infty} (I - A)^k$ \cite{Poly}. As a result of this series being convergent, higher powers of $(I-A)^k$ provide a smaller contribution to $A^{-1}$. Many matrices from Partial Differential Equations, Lattice QCD, and other applications display a significant decay in the elements of $(I - A)^k$ for larger values of $k$, further motivating the idea of probing \cite{Bali:2009,Probing, Frommer:2021}. In LQCD, in particular, a basic form of probing was first used in \cite{Wilcox:1999ab} and has become more popular with the name dilution since \cite{Foley:2005ac}.

The CP (classical probing) method is not used to directly approximate $A^{-1}$, but instead to locate its largest elements using graph coloring. Based on the decay principle above and since $(I - A)^k$ and $A^k$ have the same adjacency matrix, it is the first few powers of $A^k$ that contribute to the largest elements of $A^{-1}$. Note that the neighborhood of a node $x$ in the graph of $A^k$ is the same as the distance-$k$ neighborhood of $x$ in the graph of $A$. Therefore, the computation of $A^k$ can be avoided by working directly on the graph of $A$.

Assume that we have computed a distance-$k$ coloring of the graph of $A$ which results in $m$ colors. Conceptually, if we permuted the nodes with the same color together, $A^k$ would have $m$ color-blocks along the diagonal that are diagonal matrices. We construct the following structurally orthogonal probing vectors $z_j$, $j = 1, 2,...m$,
\begin{equation}
    z_j(i) = \begin{cases} 1 &\mbox{if } \text{color}(i) = j \\
0 & \mbox{otherwise }\end{cases}.
\label{eq:ProbingVecs}
\end{equation}
Notice that these vectors can recover exactly the trace $\text{\textbf{Tr}}(A^k) = \sum_{j=1}^{m} z_j^TA^{k}z_j$, because they completely annihilate all matrix elements outside the color-blocks along the diagonal of $A^k$ and because the color-blocks are diagonal matrices themselves. Although these diagonal blocks are dense matrices in $A^{-1}$, using these $z_j$ in the trace estimator (\ref{eq:Hutchinson}) has the same effect of annihilating all off-diagonal blocks of $A^{-1}$, or equivalently, any neighbor at distance up to $k$ from {\em any} node in the same color group. Then the accuracy of the trace estimation is the summation of the variances \cref{eq:HutchinsonVar} of the diagonal color-blocks. 

\begin{figure}[ht] 
  \centering
  \begin{subfigure}[b]{0.45\linewidth}
    \centering
    \includegraphics[width=\linewidth]{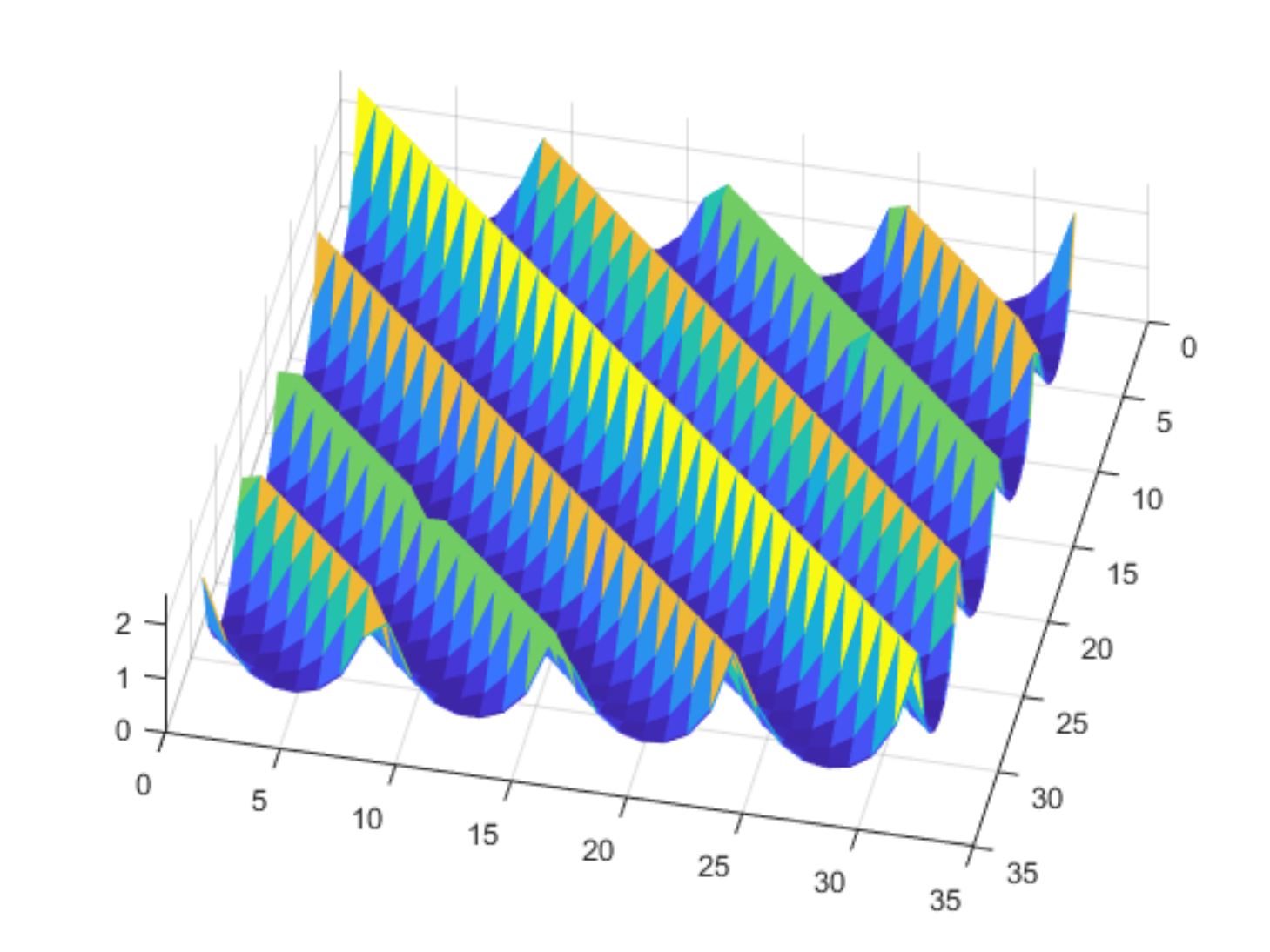} 
    \caption{$A^{-1}$ permuted into color-blocks} 
    \label{fig:permutedAinv} 
  \end{subfigure}
  \begin{subfigure}[b]{0.45\linewidth}
    \centering
    \includegraphics[width=\linewidth]{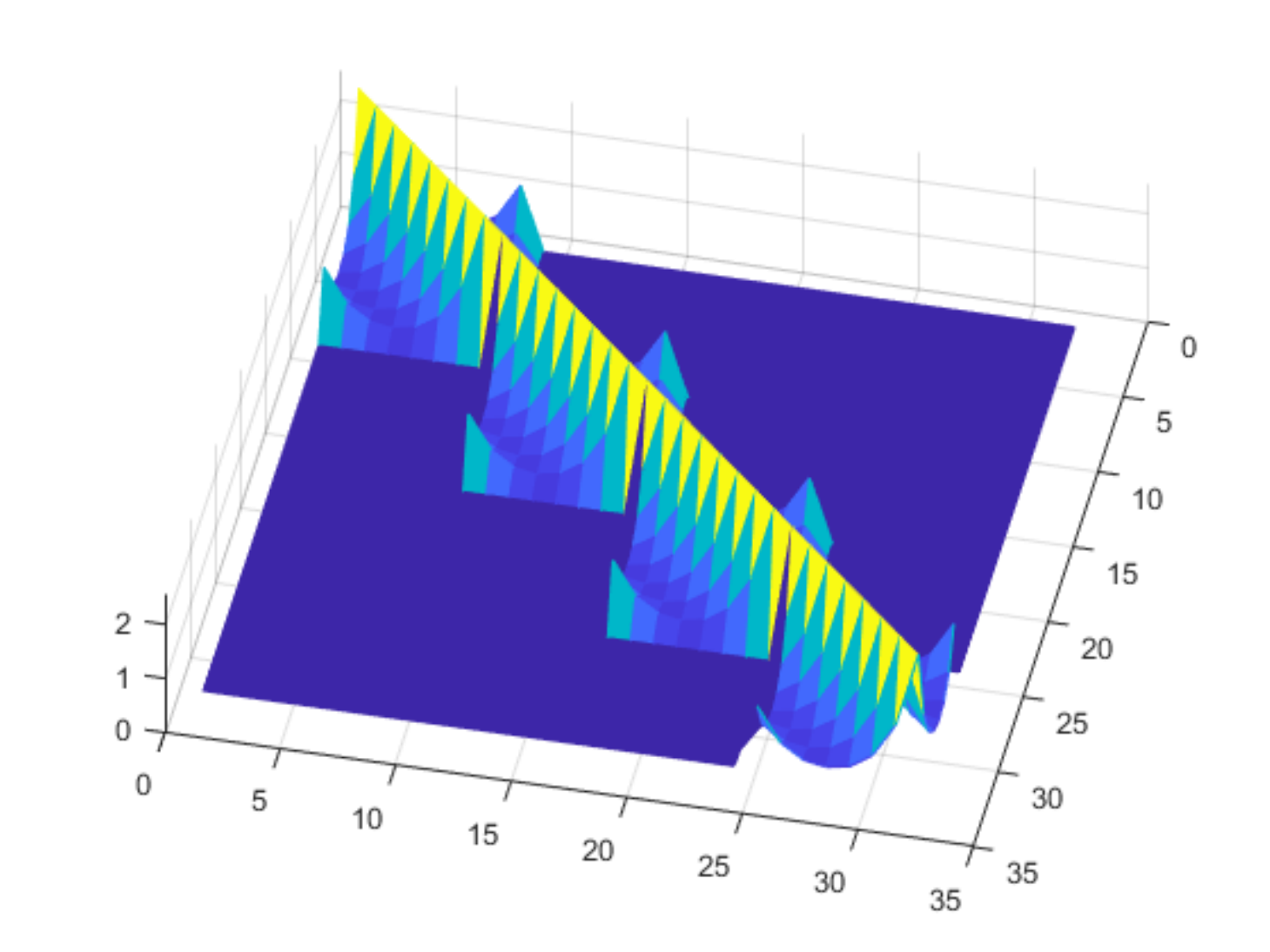}
    \caption{Permuted $A^{-1}$ after probing} 
    \label{fig:permutedAinv_odot_HHT} 
  \end{subfigure} 
  \caption{Using a shifted 1D Laplacian $A$ with 32 nodes and periodic boundary conditions, \Cref{fig:permutedAinv} shows $A^{-1}$ permuted into color-blocks based on a distance-3 coloring before probing vectors are applied. \cref{fig:permutedAinv_odot_HHT} shows the result of these color-blocks after the probing vectors are applied to the shifted Laplacian inverse, $A^{-1} \odot HH^T$.}
  \label{fig:ClassicalProbing} 
\end{figure}

\Cref{fig:ClassicalProbing} is used to display this effect. Let $A$ be a 32-node 1D Laplacian matrix with periodic boundary conditions shifted by its smallest non-zero eigenvalue so it becomes non-singular. A distance-3 coloring of this matrix yields 4 colors. Consider the permutation vector \textit{perm}  that lists the indices of all nodes in order of their color label, i.e., nodes with color 1 come first, followed by color 2, 3, and 4. Plotting $A^{-1}$ symmetrically permuted by \textit{perm} shows the color blocks along the diagonals (\cref{fig:permutedAinv}). 
\cref{fig:permutedAinv_odot_HHT} shows $A^{-1} \odot HH^T$ permuted the same way, where $\odot$ refers to the Hadamard product between two matrices, and the columns of $H$ consist of the four probing vectors from \Cref{eq:ProbingVecs}. It can be seen that every element outside the color-blocks along the main diagonal gets annihilated. 
 
Computationally, we can use a greedy coloring algorithm which takes linear time  with respect to $A^k$ and for most matrices with regular sparsity patterns provides close to optimal number of colors. The bulk of the computation is spent on the iterative method that solves for the $m$ linear systems $A^{-1}z_j$.

CP is a deterministic method. Many applications, such as LQCD, require an unbiased trace estimator (unless the deterministic accuracy can be guaranteed to be well below the statistical significance of the simulation). Moreover, if the probing vectors from the distance-$k$ coloring do not provide sufficient accuracy, we seek ways to either use $A^{-1} \odot HH^T$ as the matrix of the statistical estimator  \cref{eq:Hutchinson} or to extend CP to higher distances. In either case, the work spent on solving $A^{-1}z_j$ should be re-used and not discarded. This has been explored in \cite{HP, EHP} as described next.

\subsection{Removing Deterministic Bias}
\label{subsec:DetBias}
We note that the vectors in \cref{eq:ProbingVecs} consist of 0's and 1's in the positions determined by the colors. To remove the deterministic bias from the CP estimation, we can introduce random noise to the vectors $z_j$ similarly to one step of Hutchinson ($s=1$).
Consider the noise vector $z_0 \in \mathbb{Z}^N_2$ and apply a Hadamard product between $z_0$ and each of the probing vectors $z_j$, $j = 1, ..., m$,
\begin{equation}
    V = [z_0 \odot z_1, z_0 \odot z_2, ..., z_0 \odot z_m].
    \label{eq:noDetBias}
\end{equation}
As shown in \cite{HP}, $VV^T = HH^T$ have the same non-zero pattern, but using the vectors $v_j$ in  \cref{eq:Hutchinson} imparts no deterministic bias.

Moreover, given a sequence of random vectors, $z_0^{(i)}, i=1,\ldots , s$, we can construct the vector sets $V^{(1)}, \ldots , V^{(s)}$ as above. Using these $s\times m$ vectors in \cref{eq:Hutchinson} is the same as performing $s$ steps of Hutchinson on the variance reduced matrix $A^{-1} \odot HH^T$.

\subsection{Hierarchical Probing}
Instead of applying the CP method for a fixed distance $k$ followed by the Hutchinson stochastic estimator, it is more beneficial to continue with probing to higher distances as long as the elements of $A^{-1}$ continue to display strong decay and as long as previous work can still be reused. 

Saving computations by reusing previous work is the goal of Hierarchical probing (HP) which was initially proposed for matrices with lattice-type structure \cite{HP} and was later extended to arbitrary sparsity patterns \cite{EHP}.
The idea is to enforce a hierarchical coloring which ensures that probing vectors for smaller distance colorings are contained in the subspace of the vectors generated for larger distances, with all distances being a power of two. Therefore the trace estimation reuses the already computed quadratures $z_j^TA^{-1}z_j$ and augments them with those from higher distances.

On lattices, we can generate a hierarchical coloring by recursively partitioning a $d$-dimensional lattice into $2^d$ sub-lattices, each receiving a different color. The non-overlapping sub-lattices guarantee that if two nodes share a color at distance $k$, they must also share a color at any smaller distance, and if two nodes do not share a color at distance $k$, they will not share a color at higher distances. Each recursion step doubles the distance between nodes of the same color. The recursion stops when all nodes are given a separate color or when the requested distance is reached. A red-black coloring between recursion steps allows for intermediate colorings as the number of colors increases by a factor of $2^d$ at each recursion. 

Instead of using \cref{eq:ProbingVecs}, 
probing vectors for the HP can be generated efficiently as special permutations of the rows and columns of the Hadamard or Fourier matrices. The nested coloring implies a nesting of the subspaces of the probing vectors which can be used incrementally until the desired accuracy is achieved. 
Used in its unbiased form of \cref{eq:noDetBias} with $s=1$, this method proved particularly flexible and effective in real world LQCD problems
\cite{PhysRevD.92.031501,PhysRevD.95.114502}.

HP was extended to arbitrary lattice sizes and in particularly general sparse matrices in \cite{EHP}. These techniques can also be used with the algorithm of this paper if a hierarchical coloring is desired. However, because the number of colors required increases by a factor of 3-4 over the non-hierarchical version, we assume that users can choose a priori the required distance. 

\subsection{Deflation}
\label{sec:deflation}
A different way to reduce the variance of the estimator is to deflate the lowest singular triplets of $A$ \cite{Deflation}. 
Given $U$ and $V$ a number of approximate left and right singular vectors of the smallest singular values of $A$, we can form the oblique projector $Q = AV(U^TAV)^{-1}U^T$ and split the trace computation into two parts,
\begin{equation}
    \text{\textbf{Tr}}(A^{-1}) = \text{\textbf{Tr}}(A^{-1}Q) + \text{\textbf{Tr}}(A^{-1}(I-Q)).
\end{equation}
Since $\text{\textbf{Tr}}(A^{-1}Q)$ is easily computed as the trace of the small matrix  $(U^TAV)^{-1}U^TV$, we can apply the stochastic estimator on the $\text{\textbf{Tr}}(A^{-1}(I-Q))$ which is expected to have smaller variance.
The number of singular vectors needed to provide a significant variance reduction of the estimator is dependent on the spectral decay of the matrix $A$ and its size. For large decay or small matrix size the singular space can be computed using an iterative SVD solver on $A$ with a multigrid as preconditioner \cite{Deflation}, or using a multigrid eigensolver directly \cite{Frommer_mgeig_21}. Still, for large matrix sizes the cost of computing and applying a large number of singular vectors becomes significant. Because the goal is to approximate $A^{-1}(I-Q)$ in Frobenious norm, the accuracy of individual vectors is less important. 
In \cite{DeflationMG} we showed that hundreds or thousands of singular vectors from the coarse grid operator of multigrid can be computed efficiently and applied effectively for deflation.

Deflation works complementary to probing.
While probing effectively captures heavy elements of $A^{-1}$ occurring within some distance $k$ between nodes, deflation captures heavy connections between elements at long range distances. Therefore combining the two techniques has shown significant improvements over using one of these methods individually. As deflation does not depend on permutations, we use the same deflation as in \cite{DeflationMG} and focus solely on the effects of our new probing method.

\section{Probing for Permutations}
\label{sec:ProbwDis}

Consider the problem of finding the trace of $PA^{-1}$ where $P$ is a permutation matrix. The problem arises in Lattice QCD where $P$ corresponds to one or more displacements in the lattice. We will study this problem shortly, but let us first consider the problem for a general $P$. 

The question is how to achieve the probing goals for $PA^{-1}$.
The CP method would take powers of the matrix $AP^T$ which does not relate to how information propagates through powers of $A$ to generate $A^{-1}$. In other words, this method may not capture the largest elements of $A^{-1}$ which are at close graph distances for each node, and thus does not satisfy the design goal of probing. Moreover, when $AP^T$ has a non symmetric sparsity structure, the graph coloring problem is not well defined, although this problem can be avoided by coloring the graph of the symmetric part of a matrix.
Finally, for our LQCD application on regular lattices, the powers $(AP^T)^k$ are much denser than the corresponding $A^k$ which means a larger number of colors and thus probing vectors.

The solution is conceptually simple. Since $PA^{-1} = P\sum_{k=0}^{\infty}(I-A)^k$, we can first take powers of the matrix $A$, permute them, and then find the coloring on the associated graph of $PA^k$, or rather its symmetric part $PA^k+(PA^k)^T$.  Despite its simplicity, when this method is applied to toroidal lattices stemming from our LQCD application it creates connectivity patterns that our HP method cannot handle. However, these patterns allow for a CP-based algorithm specifically tailored for this application. 

\begin{figure}[ht] 
  \begin{subfigure}[b]{0.248\linewidth}
    \centering
    \includegraphics[width=\linewidth]{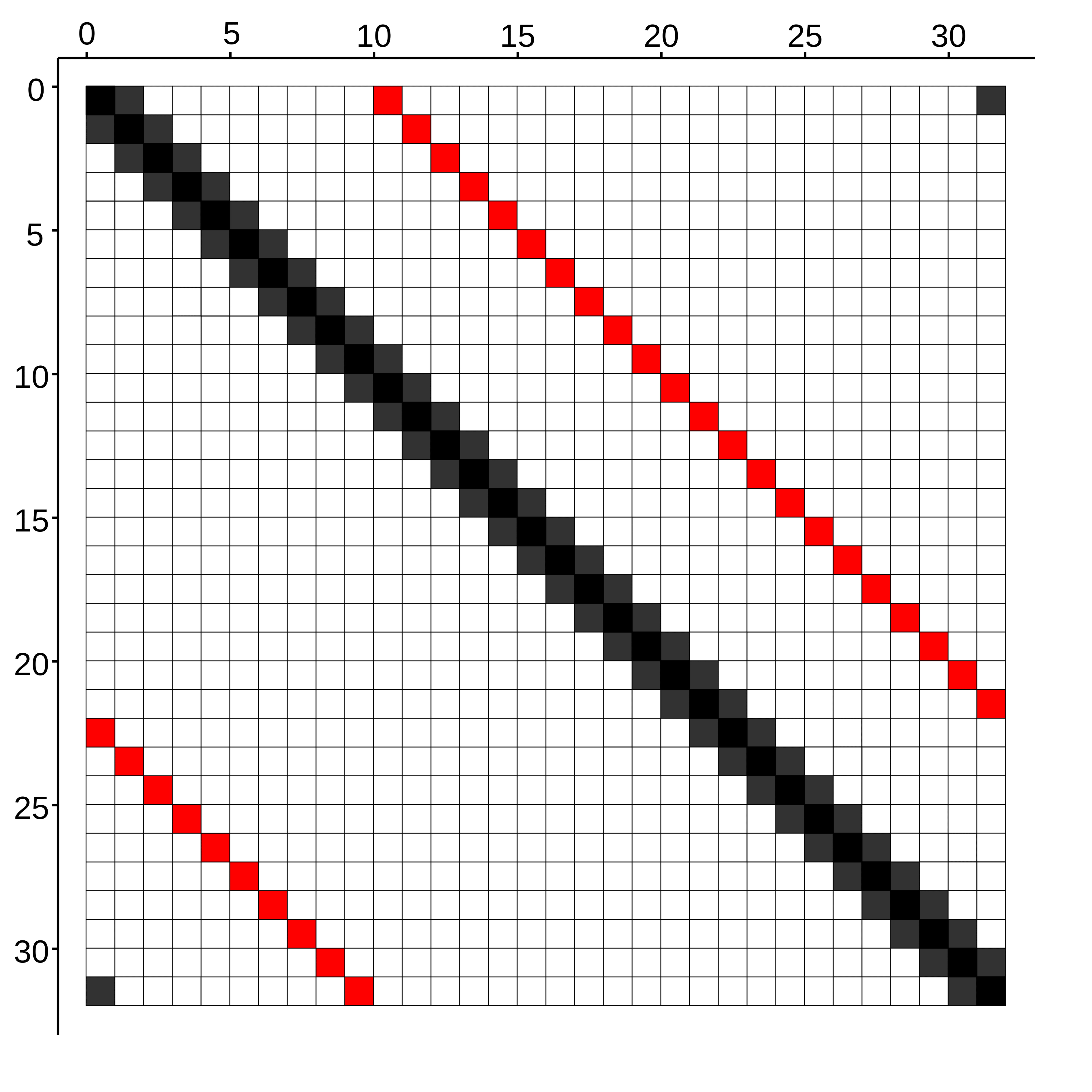} \caption{\footnotesize{Matrix $A$, 1D torus}}
    \label{fig:DisplaceMatrixa}
  \end{subfigure}
  \begin{subfigure}[b]{0.248\linewidth}
    \centering
    \includegraphics[width=\linewidth]{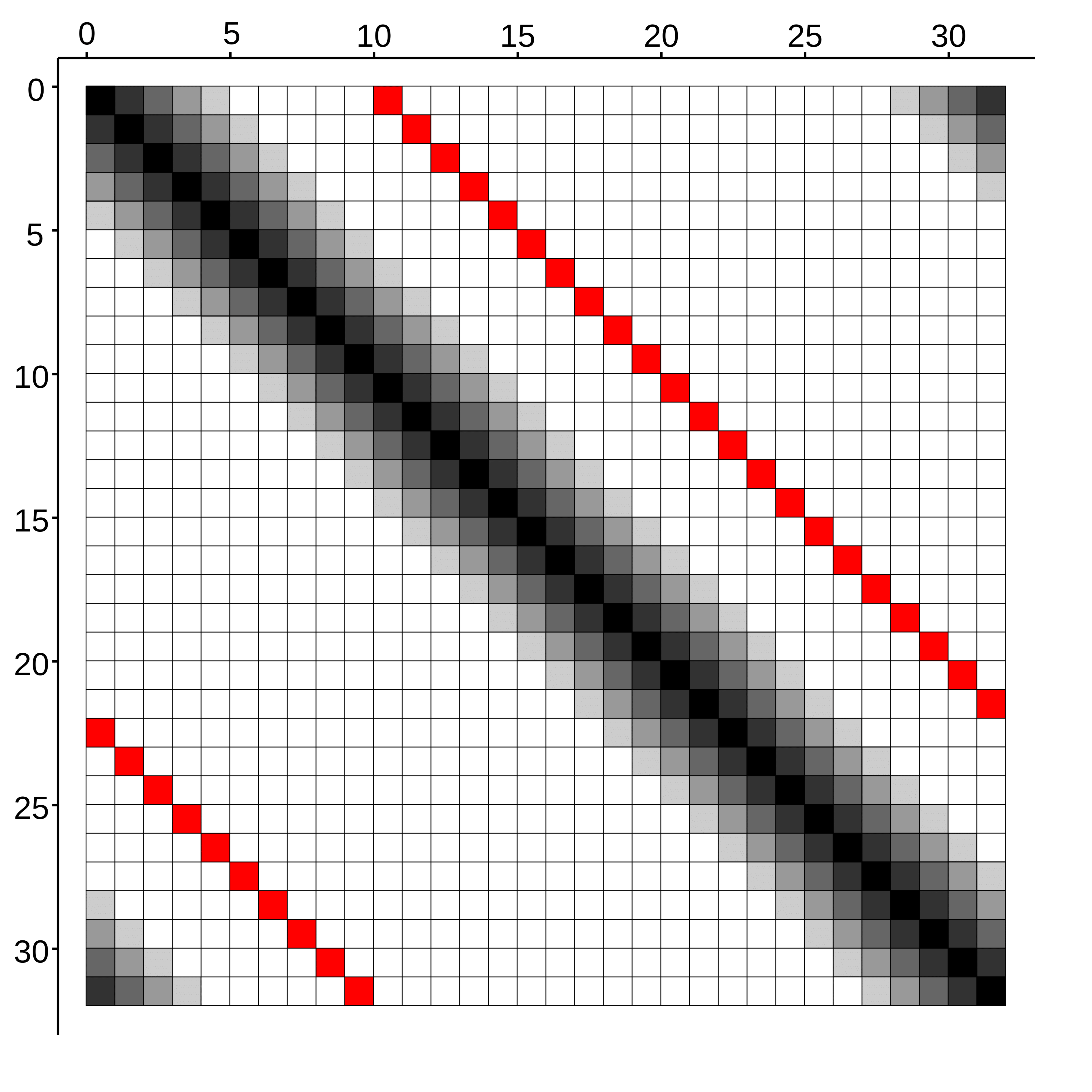}
    \caption{\footnotesize Matrix of $A^4$ }
    \label{fig:DisplaceMatrixb}
  \end{subfigure} 
  \begin{subfigure}[b]{0.248\linewidth}
    \centering
    \includegraphics[trim=1cm 0cm 0cm 0cm, clip,width=\linewidth]{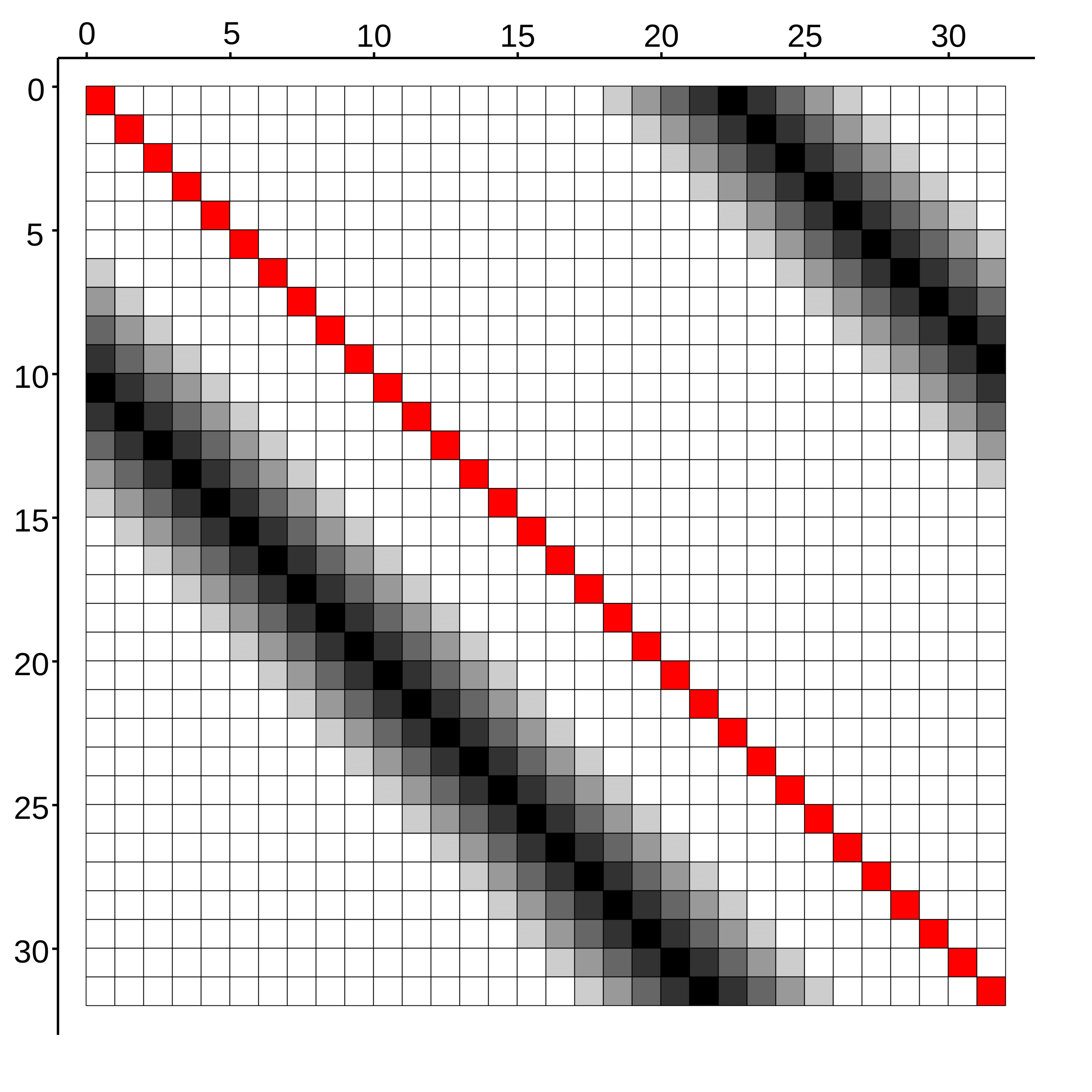} 
    \caption{\footnotesize Displace by 10 }
    \label{fig:DisplaceMatrixc}
  \end{subfigure}
  \begin{subfigure}[b]{0.248\linewidth}
    \centering
    \includegraphics[width=\linewidth]{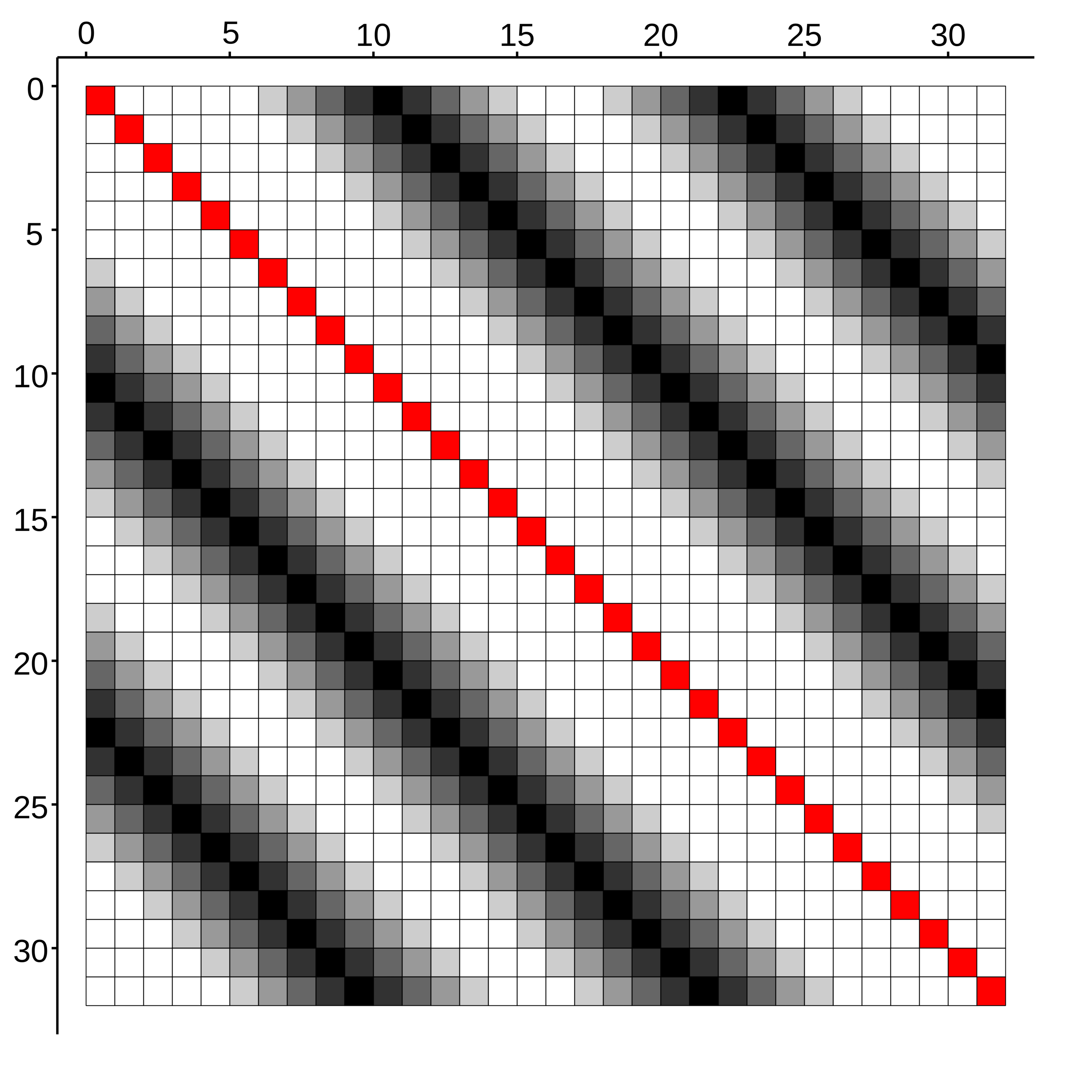} 
    \caption{\footnotesize Symmetrized }
    \label{fig:DisplaceMatrixd}
  \end{subfigure} 
  \caption{Applying a power and displacement to a matrix representation of a 1D toroidal lattice. The red diagonal represents the locations of the elements corresponding to the wanted displacement.}
  \label{fig:DisplaceMatrix} 
\end{figure}

In LQCD, the application of disconnected diagrams requires the trace of a certain projected operator which for the purpose of this discussion can be abstracted as the sum of all the elements of $A^{-1}$ that correspond to a displacement $p \in \mathbb{Z}_+^{d}$, i.e., 
$\sum_{x} A^{-1}_{ij},$
where $i$ is the index of the lattice node $x=[x_1,\ldots ,x_d]$ and $j$ is the index of node $x+[p_1, \ldots ,p_d]$. 
Let $P$ be the permutation matrix that places the required off-diagonal elements onto the main diagonal. The corresponding permutation index is computed in MATLAB as
\begin{center}
{\tt perm = Coord2Index(mod(Index2Coord([1:N],D)+p,D), D);}
\end{center}
where the two functions are the maps between lattice coordinates and the particular index ordering of the application.
The inverse permutation $P^T$ simply maps a lattice point $y$ to $y-[p_1,\ldots ,p_d]$.
The idea of coloring the graph of $PA^k+(PA^k)^T$ is shown in \Cref{fig:DisplaceMatrix} for the matrix of a 1D periodic lattice of 32 points with $p=10$ and $k = 4$. The red locations indicate the required diagonal at displacement $p=10$. The gray scale diagonals show the magnitude of the non-zeros of the following matrices: \Cref{fig:DisplaceMatrixa} of the matrix $A$; \cref{fig:DisplaceMatrixb} of the $A^4$; \Cref{fig:DisplaceMatrixc} of the matrix $PA^4$, and  \Cref{fig:DisplaceMatrixd} of the matrix $PA^4+(PA^4)^T$. Based on the aforementioned decay property, coloring this last matrix would eliminate the heaviest elements of the inverse.

As with CP we use a greedy linear time algorithm to color $PA^k+(PA^k)^T$. However, by working directly on the lattice we are able to speed up the distance-$k$ coloring process. Given a node $x$ with lattice coordinates $[x_1, .., x_d]$, we do not find the distance-$k$ neighborhood of $x$, but rather the distance-$k$ neighborhoods centered at 
\begin{equation}
x^+ = [x_1+p_1,\ldots , x_d+p_d] \mbox{ and } x^- = [x_1-p_1,\ldots ,x_d-p_d].
\label{eq:x+x-}
\end{equation}
Displacements in both $p$ and $-p$  directions enforce a symmetric matrix structure. We denote the distance-$k$ neighborhood of $x$ for displacement $p$ as, 
\begin{align}
\begin{split}
  N^d(x, p, k) & = N^d(x^+, 0, k) \cup N^d(x^-, 0, k)\\
  &= \{y : \|y-x^+\|_1 \leq k \} \cup
     \{y : \|y-x^-\|_1 \leq k \}.
\end{split}
  \label{eq:Neighborhood}
\end{align}
During coloring, we exclude $\{x\}$ from the neighborhood, and when the dimension $d$ is implied, we omit the superscript.

We make three observations. First, the main diagonal of the original $A^{-1}$, whose elements are typically of the largest magnitude, is part of the off-diagonal structure of $PA^{-1}$ and contributes to the estimator variance. However, the $(x,x)$ elements of this diagonal are now displaced to the $(x,x^-)$ lattice points in the $N^d(x, p, 0)$, so our new method eliminates them immediately for any probing distance.
Equivalently, because of the assumed decay, the elements of next-highest magnitude in $A^{-1}$ will be in the diagonals closest to the main or at distance $k=1$ from it. The decay continues with higher distances $k$. Therefore the new algorithm includes in the neighborhoods $N^d(x, p, k)$ all original distance-$k$ neighbors of the points $x^+$ and $x^-$ as these will have the largest weight. Finally, we note that although $k=0$ removes the old main diagonal (the graph of $P+P^T$), in practice probing is meaningful for $k\geq 1$.

\subsection{Coloring with Displacements Algorithm}
\label{subsec:DispAlg}

Once we have defined the neighborhood of each node in the displacement graph we can use a simple greedy approach to color it \cite{GreedyCol}. The number of colors translates to the number of iterations in the stochastic estimator. It is not as critical to minimize this number as more vectors/iterations could imply a larger variance reduction. However, this additional reduction beyond the best distance-$k$ coloring is hard to quantify and may not be more effective than using extra random noise vectors. 
The order in which nodes are visited by the greedy algorithm is thus important.

We have experimented with some common visitation orders such as natural and red-black orderings, a completely random order, and a random red-black where the order of the nodes within a color is random. In addition, we tested a domain decomposition idea, where an independent set of the graph of $A^i$ was constructed for various $i$'s, and then breadth first search was used to add neighborhoods to each of these centers (for $i=1$ this reverts to red-black). 
After extensive testing we observed that in most cases, natural and red-black orders achieved the least amount of colors. Surprisingly, thousands of runs of the random variants yielded only marginal improvements, and the domain decomposition idea deteriorated with increasing $i$. We believe this is due to the well-structured connections of the lattice.

\Cref{alg:DisColoring} shows how to work directly on the lattice $\mathbb{Z}^d_D$ to apply the greedy distance-$k$ coloring algorithm for a displacement vector $p$, and for a user-defined visitation order.
It returns a vector {\tt Colors} which can be used in \cref{eq:ProbingVecs} to generate the probing vectors. 
To avoid re-computing the neighborhood for each lattice point, \Cref{alg:CreateStencil} builds first a ``stencil'' of coordinate offsets that when added to the coordinates of some point $x$ returns the coordinates of the points in $N(x,p,k)$.
Because every lattice node is of the same degree, it is clear that the maximum number of colors produced by the greedy algorithm is one more than the degree of a node, i.e., colors are less or equal to $|N(x,p,k)|+1 =  len(Stencil(:, 1))+1$. A bit array of this size can be used to record the colors used for each neighborhood and find the first color not in use.
The colors returned by \Cref{alg:DisColoring} are used in \cref{eq:ProbingVecs} and then \cref{eq:noDetBias} to generate the unbiased probing vectors to be applied on the displaced inverse $PA^{-1}$.

\begin{figure*}
\input{Algorithms/ColoringAlg}
\mbox{}\\ \vspace{-30pt}
\input{Algorithms/StencilAlg}
\end{figure*}

The size of the distance-$k$ $L_1$ ball on the lattice is $O(k^d)$ and the stencil contains two such balls in $N(x,p,k)$. To union the two stencil balls we have to remove duplicates when the balls overlap, which can be obtained by sorting the elements. This gives a complexity $O(k^dd \log k)$ to generate the stencil. The dominant part of the complexity is the linear time greedy algorithm which visits the neighborhoods $N(x,p,k)$ for each $x$, and therefore the algorithm's complexity is $O(Nk^d)$.

Although the algorithm we presented is for any $d$-dimensional displacement, in practical LQCD problems the  displacement occurs only in the z space-time direction. For convenience our theoretical discussion considers the displacement to be in the 1st dimension, i.e., $p = p_1$ and $p_2=\ldots =p_d = 0$.




\subsection{Lower Bound on the Number of Colors}
\label{subsec:LowerBound}

The chromatic number of a graph must be at least the size of its maximal clique. 
In our problem, the neighborhood of every lattice node is the union of two $L_1$ balls so we seek to identify its maximal clique. This is complicated by the wrap-around property of the torus which adds additional constraints to the coloring and thus the results depend not only on $p$ and $k$, but also on the size $D_i$ of each dimension. 
To avoid this complication, we ignore the toroidal property which, for sufficiently large $D_i$, is equivalent to considering the lattice $\mathbb{Z}_{\infty}^d$ which is infinite in all $d$ dimensions. By removing these constraints from the coloring algorithm, the size of the maximal clique of the infinite lattice may be smaller, and thus its size will still be a lower bound to the chromatic number of the finite toroidal lattice.
We call the number of colors required to distance-$k$ color the infinite lattice with displacement $p$, $col(\mathbb{Z}^d_{\infty},p,k)$.

Without displacement, $p=0$, each neighborhood $N(x,0,k)$ is an $L_1$ ball of radius $k$. Any two points in this ball are at $L_1$ distance $2k$ or less. Therefore, the maximal clique of the distance-$k$ graph of $N(x,0,k)$ should be the nodes inside the $L_1$ ball of radius $\lfloor \frac{k}{2} \rfloor$. If $k$ is odd, this $L_1$ ball is extended by one point in one dimension. The lower bound on the chromatic number is given by the size of this clique
\begin{equation}
    col(\mathbb{Z}^d_{\infty},0,k)= \left\{
\begin{array}{ll}
 |N^d(\mathbf{0}, 0, \frac{k}{2})| &  \mbox{if $k$ is even}\\
 |N^d(\mathbf{0}, 0, \lfloor \frac{k}{2} \rfloor)| + |N^{d-1}(\mathbf{0}, 0, \lfloor\frac{k}{2} \rfloor)|  &  \mbox{if $k$ is odd}   \\
\end{array}
\right. ,
\label{eq:p=0Bounds}
\end{equation}
where $\mathbf{0}=[0, ..., 0]$ is chosen as a representative neighborhood center.
Recurrence relations can be derived to compute this number for any dimension, although general closed forms for an arbitrary number of dimensions are not known. More details can be found in \cite{ComputingContinuousDiscretely, HP, EHP}.

With displacement ($p>0$), the $L_1$ balls of a neighborhood $N(x,p,k)$ are not centered around the node $x$, resulting in different coloring patterns. We characterize the number of colors needed, first for $p\geq k$ and then for $p<k$. Proofs are given in Appendix A. 

\begin{theorem}
Let $x \in \mathbb{Z}^d_{\infty}$. If $p \geq k $, then $\forall y \not = x$ with $y_1 = x_1$, it holds $y \notin N(x, p, k)$. 
\label{thm:1}
\end{theorem}

\begin{figure}[htb]
    \centering
      \begin{subfigure}[b]{.4\linewidth}
    \centering
    \includegraphics[width=\linewidth]{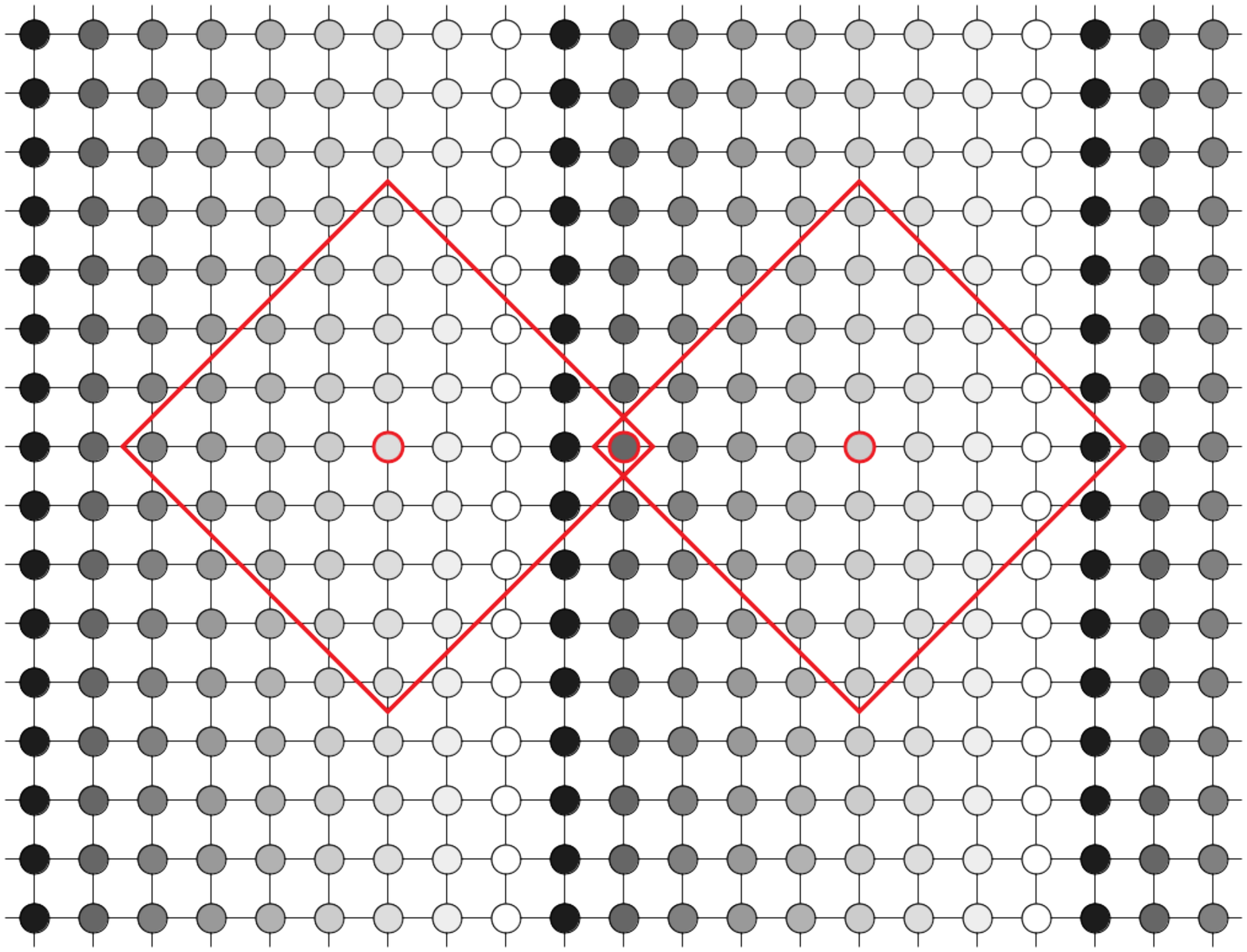} 
    \caption{$N^2(x, 4, 4)$} 
    \label{fig1:p4k4} 
  \end{subfigure}
  \begin{subfigure}[b]{.4\linewidth}
    \centering
    \includegraphics[width=\linewidth]{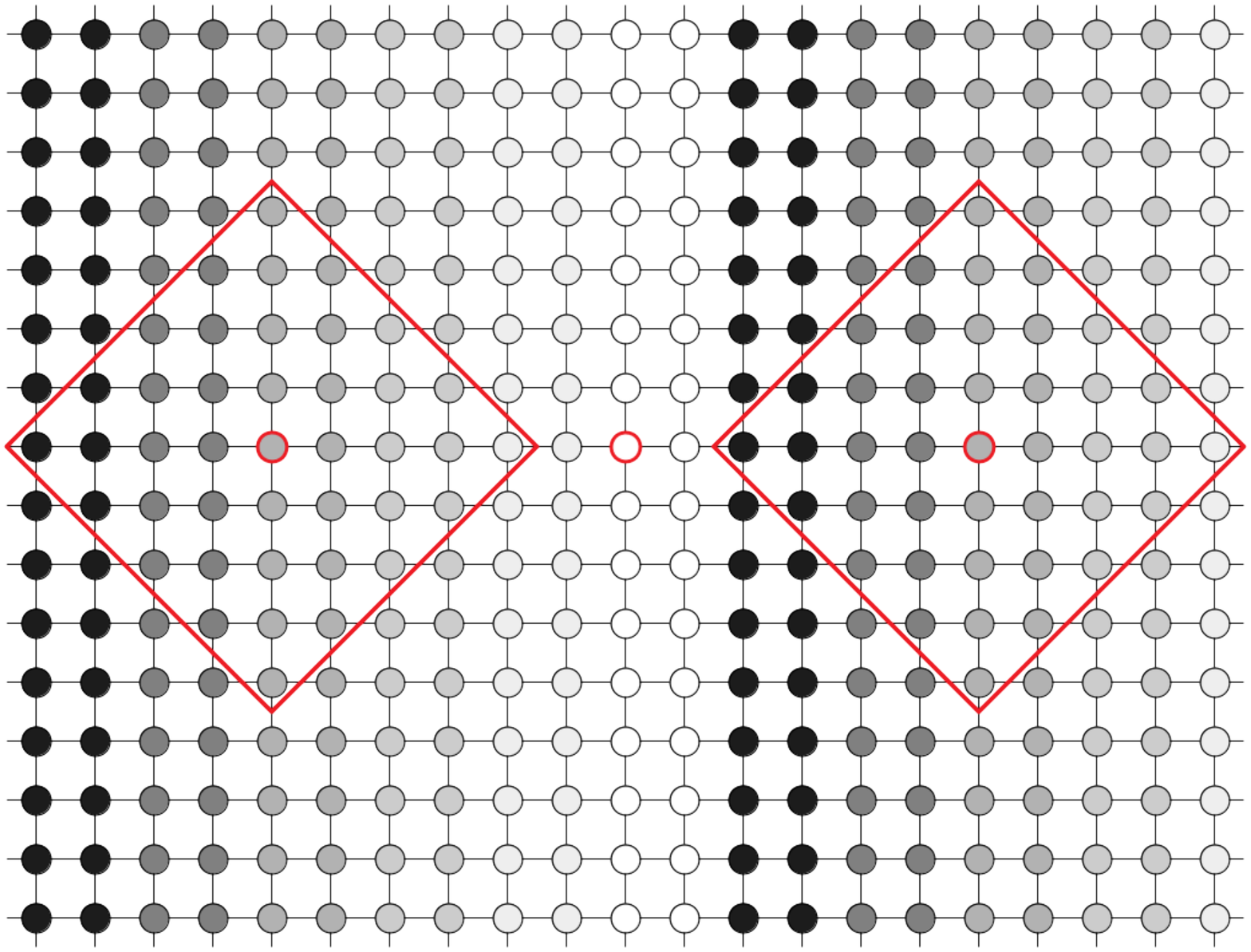} 
    \caption{$N^2(x, 6, 4)$} 
    \label{fig1:p6k4} 
  \end{subfigure} 
    \caption{The neighborhood $N^2(\mathbf{0},p,k)$ and how 1D coloring is sufficient when $p \geq k$. }
    \label{fig:p>=k}
\end{figure}

The above theorem implies that when $p\geq k$ all nodes with the same $x_1$-coordinate can share the same color, reducing the $d$-dimensional coloring problem to a 1D problem. An example of this can be seen in \Cref{fig:p>=k}.
To find the lower bound on the number of colors we consider the two sub-cases, $p = k$ and $p > k$, separately.

\begin{theorem}
If $p = k$,  then $col(\mathbb{Z}^d_{\infty},p,k) = 2k+1$.
\label{thm:2}
\end{theorem}

\begin{theorem}
If $p > k$, then  $col(\mathbb{Z}^d_{\infty},p,k) = \lceil \frac{2p}{p-k} \rceil=\lceil \frac{2k}{p-k}\rceil +2$.
\label{thm:3}
\end{theorem}

\begin{figure}[htb]
    \centering
      \begin{subfigure}[b]{.3\linewidth}
    \centering
    \includegraphics[width=\linewidth]{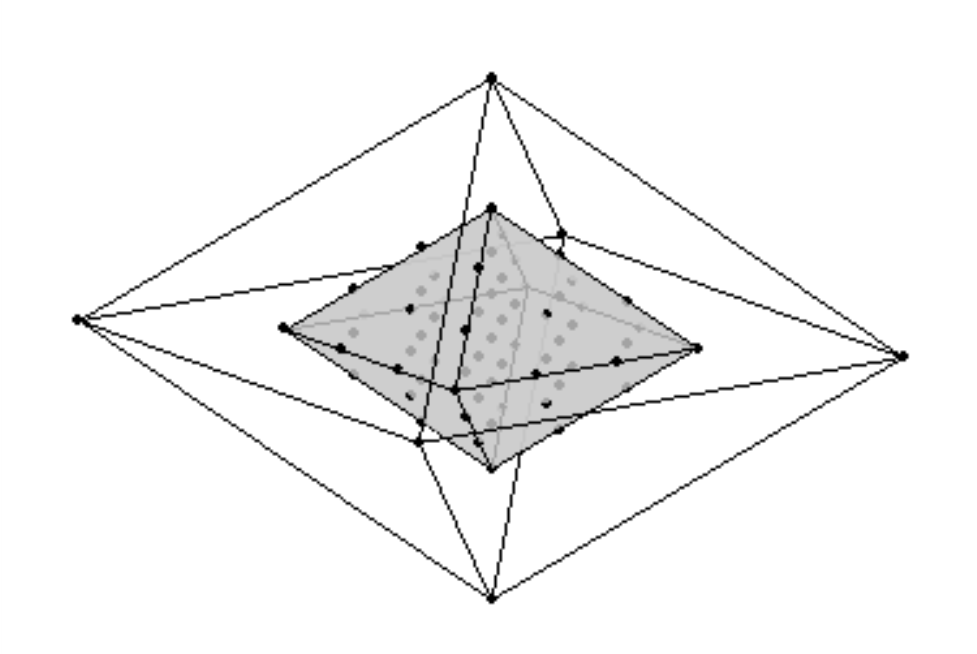} 
    \caption{$k = 6$, $p = 0$} 
    \label{fig1:Ck6p0} 
  \end{subfigure} 
  \begin{subfigure}[b]{.3\linewidth}
    \centering
    \includegraphics[width=\linewidth]{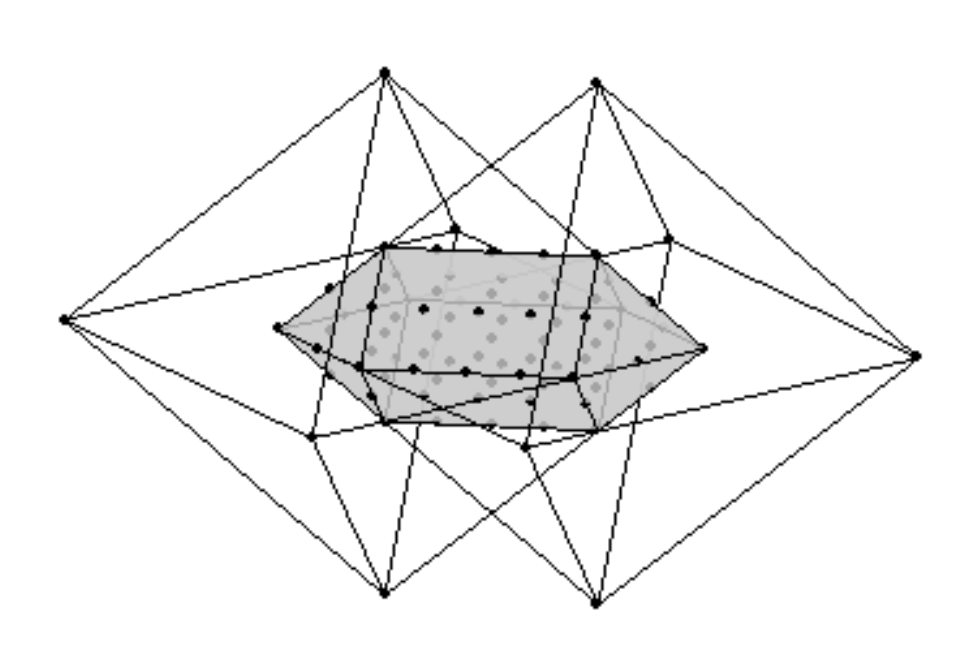} 
    \caption{$k = 6$, $p = 2$} 
    \label{fig1:Ck6p2} 
  \end{subfigure}
  \begin{subfigure}[b]{.3\linewidth}
    \centering
    \includegraphics[width=\linewidth]{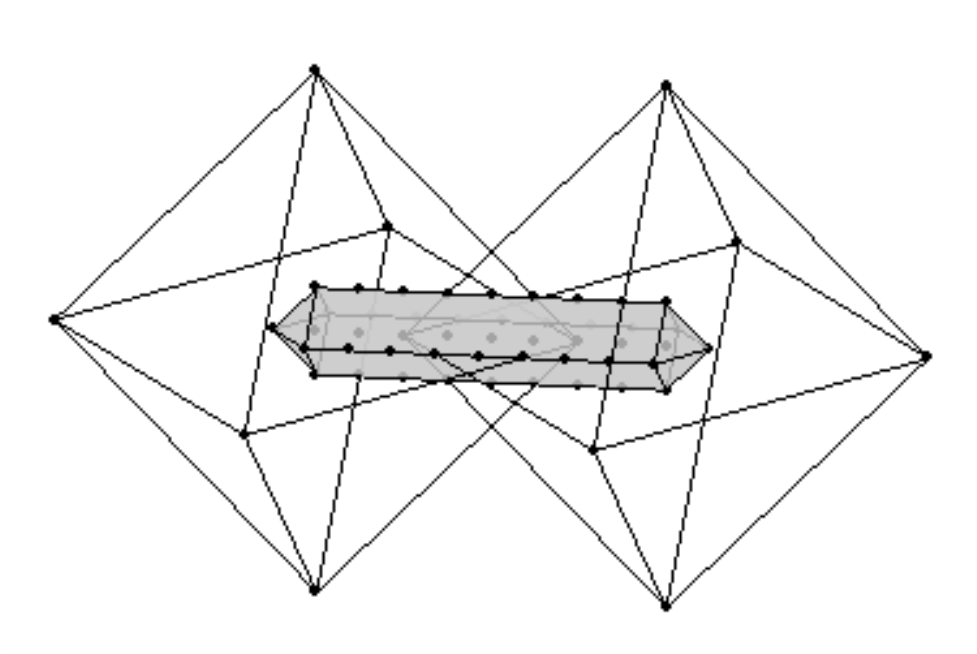} 
    \caption{$k = 6$, $p = 4$} 
    \label{fig1:Ck6p4} 
  \end{subfigure} 
    \caption{The distance-$k$ clique shown in grey of the neighborhood $N^3(\mathbf{0},p,k)$ which is shown as wire frames, when $p < k$ and $(k + p)$ is even as described in \Cref{thm:evenC}.}
    \label{fig:evenCase}
\end{figure}

When $p<k$, the two $L_1$ balls centered around $x^-$ and $x^+$ overlap. Next, we identify the maximal clique in this neighborhood for which all points are at distance $k$ or less considering displacement $p$. As before, we center the neighborhood at $x = \mathbf{0}$. The case where ($k+p$) is even is considered in \Cref{thm:evenC}, an example of which is shown in \Cref{fig:evenCase}.

\begin{theorem}
\label{thm:evenC}
Assume $(k + p)$ is even and $p<k$. Let, $\alpha = \lfloor\frac{k+p}{2}\rfloor$, $\beta = \lfloor\frac{k-p}{2}\rfloor$, and define the set 
\begin{equation}
    C(d,\alpha,\beta) = \left\{ x: \| x\|_1 \leq \alpha 
    \mbox{ and }
    \sum_{i=2}^d |x_i| \leq \beta \right\}. \label{cond:1}
\end{equation}
Then $\forall x, y \in C(d,\alpha,\beta)$, $x \in$ N(y, p, k), i.e., $C(d,\alpha,\beta)$ constitutes a distance-$k$ clique.
\end{theorem}

\begin{figure}[h]
    \centering
      \begin{subfigure}[b]{.3\linewidth}
    \centering
    \includegraphics[width=\linewidth]{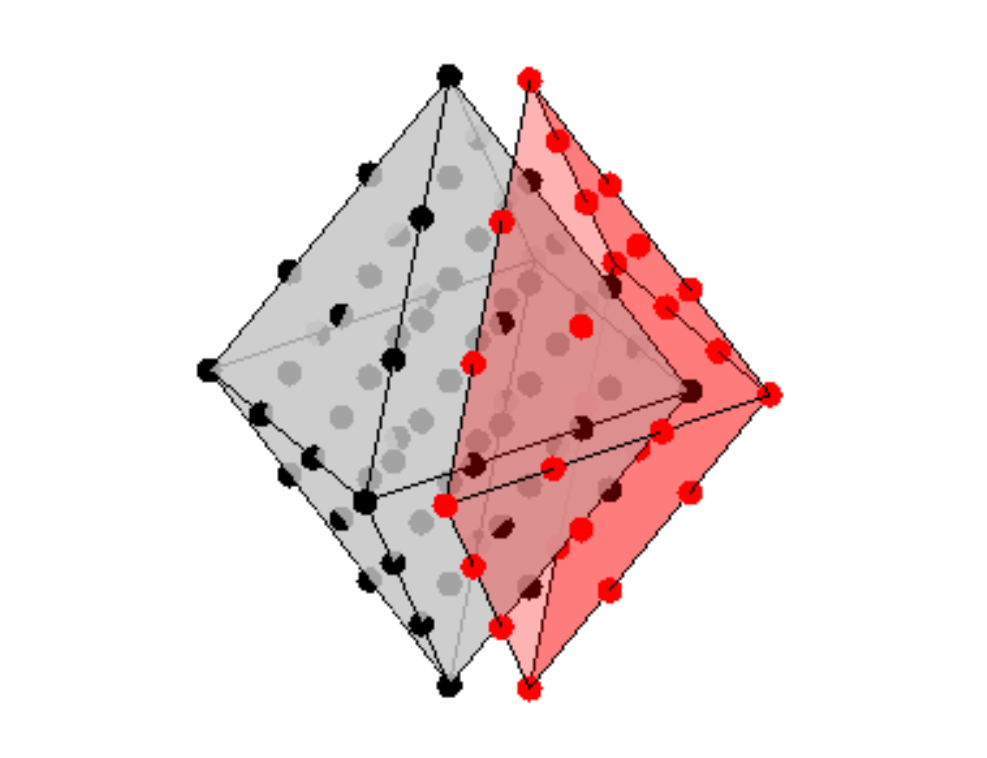} 
    \caption{$k = 7$, $p = 0$} 
    \label{fig1:Ck7p0} 
  \end{subfigure} 
  \begin{subfigure}[b]{.3\linewidth}
    \centering
    \includegraphics[width=\linewidth]{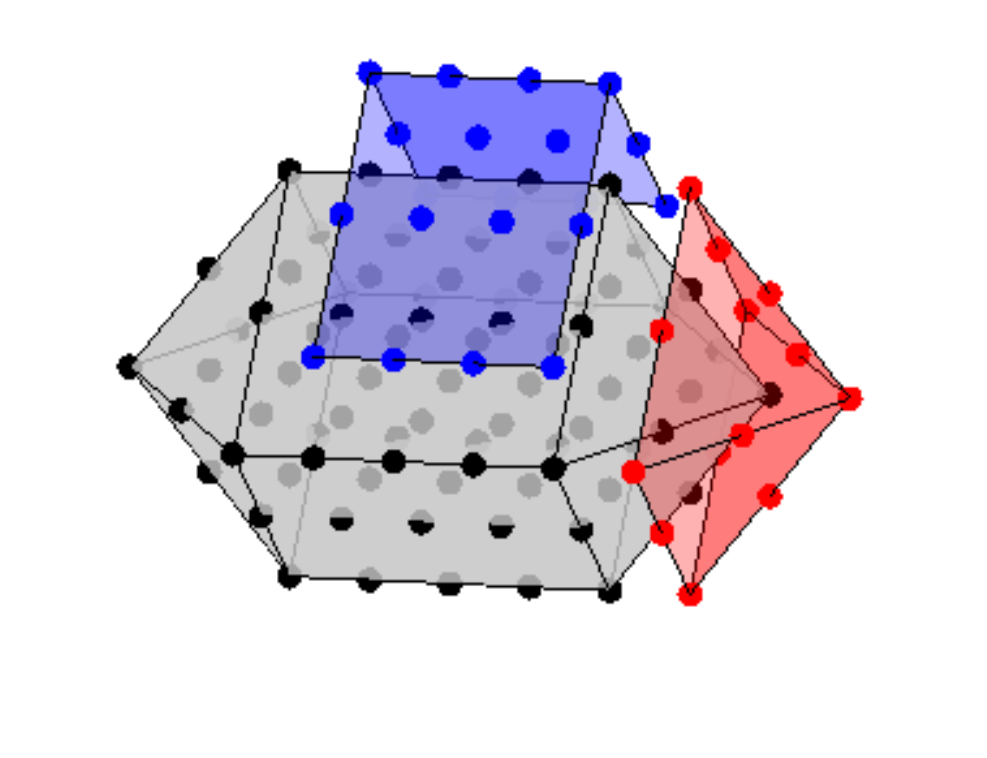} 
    \caption{$k = 7$, $p = 2$} 
    \label{fig1:Ck7p2} 
  \end{subfigure}
  \begin{subfigure}[b]{.3\linewidth}
    \centering
    \includegraphics[width=\linewidth]{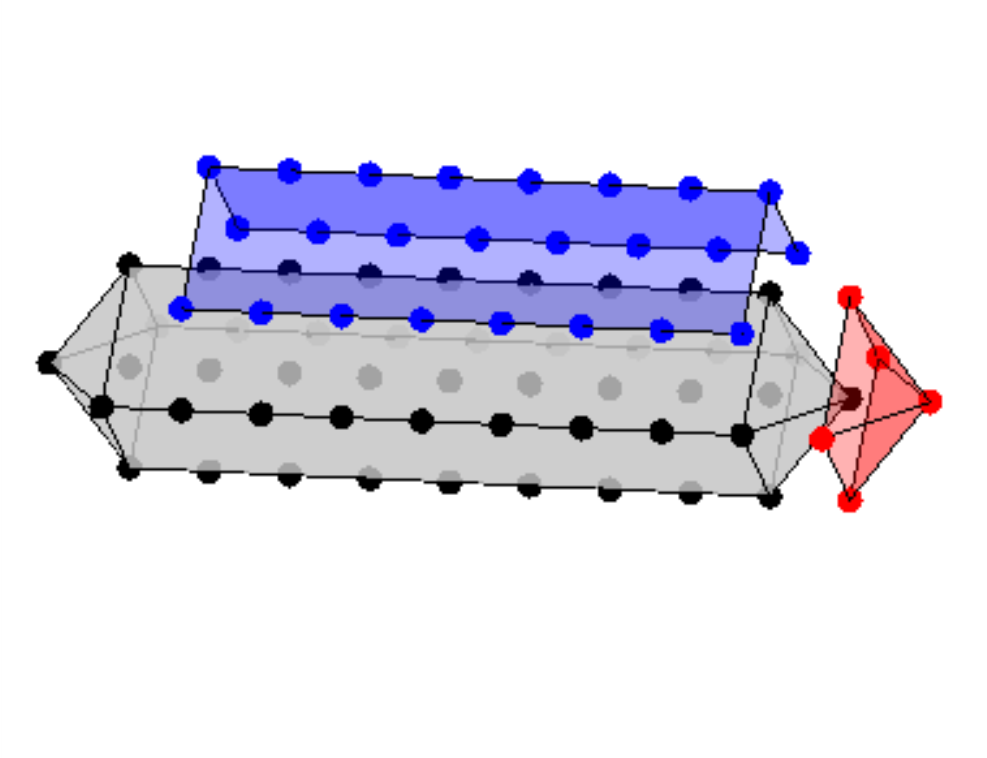} 
    \caption{$k = 7$, $p = 4$} 
    \label{fig1:Ck7p4} 
  \end{subfigure} 
    \caption{Distance-$k$ cliques of $N^3(\mathbf{0},p,k)$ when $(k + p)$ is odd as shown in \Cref{thm:oddC}. Set $C(d,\alpha,\beta)$ is the grey set in the center, set $S$ is the red hyper-surface on the right, $T$ is the blue hyper-surface on the top.}
    \label{fig:oddCase}
\end{figure}

For $(k+p)$ is odd, \cref{eq:p=0Bounds} shows that when $p=0$ the clique needs to be extended by one hyper-surface. In \Cref{thm:oddC} we prove that for $p>0$ the clique requires two additional hyper-surfaces as depicted in \Cref{fig:oddCase}.

\begin{theorem}
\label{thm:oddC}
Assume $(k + p)$ is odd and $k>p$. Define 
$C' = C(d,\alpha,\beta) \cup T \cup S$, where $C(d,\alpha,\beta)$ is defined in \cref{cond:1} and
\begin{align}
&  T =  \{x: -(p-1) \leq x_1 \leq p \mbox{ and } 1\leq x_2 < \beta+1 \mbox{ and }  \sum\nolimits_{i=2}^d |x_i| = \beta + 1\} \label{cond:2}, \\
&   S =
 \{ x: p+1 \leq x_1 \leq \alpha+1  \mbox{ and }  |x_2| \leq \beta  \mbox{ and }  \| x\|_1 = \alpha +1 \}. \label{cond:3}
\end{align}
Then $\forall x, y \in C'$, $x \in$ N(y, p, k), i.e., $C'$ constitutes a distance-$k$ clique.
\end{theorem}

Finally, to count the number of points in the clique for any combination of $d,p,k$ we can use the recursive \Cref{alg:RecMin}. \Cref{tab:Formulas} shows the analytic formulas for the size of $C(d,p,k)$ obtained by the nested summations of points over all dimensions for lattices with $d = 1, 2, 3,  4$ when $k+p$ is even and $k>p$. 
For $k+p$ odd, we need to add also the size of $d-1$ dimensional hyper-surfaces $S$ and $T$. It is not hard to see that $|S|+|T| = |C(d-1,\alpha,\beta)|$. Therefore, we arrive at the following general lower bound for the number of colors of our algorithm,
\begin{equation}
    col(\mathbb{Z}^d_{\infty},p,k)= \left\{
\begin{array}{ll}
 2k+1                               & \mbox{if $k=p$} \\
 \lceil \frac{2p}{p-k} \rceil             & \mbox{if $k<p$} \\
 |C(d,\alpha,\beta)| &  \mbox{if $k>p$, $k+p$ even}\\
 |C(d,\alpha,\beta)| + |C(d-1,\alpha,\beta)| & \mbox{if $k>p$, $k+p$ odd}   \\
\end{array}
\right. .
\label{eq:pBounds}
\end{equation}

\begin{algorithm}[h]
\setcounter{AlgoLine}{0}
\DontPrintSemicolon
  \KwInput{\\ \qquad $k$ = Coloring distance; \hspace{60pt} $p$ = Displacement (in the first dimension) \\ \qquad $s$ = The current distance traveled;  \hspace{8pt} $d$ = Current dimension level \\ 
  \qquad $min\_Colors$ = Number of colors needed so far}
  \SetKwFunction{MinNumColors}{Min\_Num\_Colors}
  \Indm\nonl\MinNumColors{$k,p,s,d,min\_Colors$}
  \Indp
  
  \If{$p > k$}{
    $min\_Colors = \lceil \frac{2\text{*}k}{p-k}\rceil + 2$;
  
    \KwRet{$min\_Colors$}
  }
  
  \If{$d == 0$}{
     $min\_Colors = min\_Colors + 1$;
     
     \KwRet{$min\_Colors$}
  }
  
  \If{$d == 1$}{
     $min\_Colors = min\_Colors + 2$*$(\lfloor \frac{k-p}{2} \rfloor - s) + 1$;
     
     \KwRet{$min\_Colors$}
  }
  
  \For{$i = -\lfloor\frac{k-p}{2}\rfloor + s$ : $\lfloor\frac{k-p}{2}\rfloor - s$}{
    $min\_Colors = $\MinNumColors{$k,p,s+\left| i\right|,d-1,min\_Colors$};
  }
  \KwRet{$min\_Colors$}

\caption{Recursive Function to Find the Lower Bound on Colors Needed}
\label{alg:RecMin}
\end{algorithm}

\begin{table}[htb]
\centering
\begin{tabular}{c||c|}
\hline
\multicolumn{1}{|l||}{d}& Size of the clique $C(d,\alpha,\beta)$ for $k>p$ and $(k + p)$ even\\ \hline \hline
\multicolumn{1}{|c||}{1} & $2\alpha + 1$\\ \hline
\multicolumn{1}{|c||}{2} & $-2\beta^2 + 4\alpha\beta +2\alpha+1$\\ \hline
\multicolumn{1}{|c||}{3} & $-\frac{8}{3}\beta^3 + (4\alpha -2)\beta^2 + (4\alpha + \frac{2}{3})\beta + 2\alpha + 1$ \\ \hline
\multicolumn{1}{|c||}{4} & $\frac{1}{3}(2(4 \beta^3 + 6\beta^2 + 8\beta +3)\alpha - 6\beta^4 - 8\beta^3 - 6 \beta^2 +2\beta +3)$ \\ \hline
\end{tabular}
\caption{Formulas for size of the clique $|C(d,\alpha,\beta)|$, if $k > p$ and $(k+p)$ is even, with $\alpha = \lfloor\frac{k+p}{2}\rfloor$ and $\beta = \lfloor\frac{k-p}{2}\rfloor$.
If $(k+p)$ is odd, use \cref{eq:pBounds}.}
\label{tab:Formulas}
\end{table}

\subsection{Clearances}
\label{subsec:Clearances}

The LQCD application of disconnected diagrams requires the computation of traces not only for one but for multiple displacements (e.g., $p = 0, \ldots , 8$). Using different colorings to individually find each of the traces is computationally prohibitive as we would have to solve a different set of linear systems for each of the nine displacements. Therefore, it is natural to ask whether the probing vectors from one displacement can be used effectively for other ones. \Cref{thm:Clearance} shows that if a distance-$k$ coloring generated for displacement $p$ is used for displacement $p+ \lambda$ or $p-\lambda$, then it clears at least distance $\max(k-\lambda,0)$. 
\begin{theorem}
\label{thm:Clearance}
$N(\mathbf{0}, p\pm \lambda, k-\lambda) \subseteq N(\mathbf{0}, p, k)$, for any $\lambda \leq k$.
\end{theorem}

Based on this theorem, a specific $(p,k)$-coloring, i.e., a distance $k$-coloring for displacement $p$, will also be effective in reducing variance for nearby displacements. However, its effectiveness declines for farther displacements. In our LQCD experiments we show that choosing larger valued $(p,k)$ pairs is more beneficial.

\subsection{Multiple Displacements}
\label{subsec:MultDis}


The diminishing clearance achieved from $(p,k)$-coloring to farther displacements motivates the idea of finding a single distance-$k$ coloring for a graph stemming from multiple displacements. The goal is to spread the effectiveness of a power $k$ to more values of $p$, instead of using one $p$ and a high $k$ value, while still using less colors than all displacements individually. 
Given a list of displacements, $p_1, p_2, 
\ldots, p_n$, the neighborhood of a node $x$ can be constructed as,
\begin{equation}
    N(x, [p_1, ..., p_n], k) = N(x, p_1, k) \cup \ldots \cup N(x, p_n, k).
\end{equation}
\cref{alg:DisColoring} can be modified to do this by calling $Create\_Stencil$ for multiple different $p$ vectors and unioning the created stencils together. 

As expected from \cref{thm:Clearance}, we observed that the resulting clique is smaller when the displacements $p_1, p_2, \ldots, p_n$ are successive. 
In fact, when the distance between each displacement is greater than $k$, this method returns a similar number of colors to the total number returned when each of the displacements are applied individually.
However, in our LQCD experiments even successive multiple displacements did not yield improvements in variance over just using one of the higher displacements (say $p_n$) with distance larger than $k$. We believe this is due to the fact that smaller displacement traces have significant higher magnitude thus requiring less variance reduction. This is discussed in the experiments section.


\subsection{Tiles}
\label{subsec:Tiles}

Despite the linear complexity of \cref{alg:DisColoring}, practical lattice sizes reach $64^4$ and often larger, and the neighborhood size is $O(k^4)$ (e.g., for $p=0, k=10$ there are 8361 neighbors to visit). It is clear therefore that we should avoid running the method every time a new trace problem is solved. One solution is to generate and save in a database colorings for most useful lattice sizes. However, the regular structure of the lattice results in coloring patterns that repeat across the lattice. This is one of the motivations for tiling: we color a smaller toroidal lattice, the tile, and repeat its coloring throughout the lattice. Small tiles can be generated at runtime, and several common larger tiles can be saved in the aforementioned database.

The second motivation comes from the effect of lattice size to the number of colors.
While our analysis was based on $\mathbb{Z}_{\infty}^d$, with a wrap-around structure the additional constraints make the number of colors sensitive to the lattice size. For example, the distance-1 coloring of a non-periodic 1D lattice requires 2 colors, while for the toroidal lattice we need 2 colors when $D_1$ is even and 3 colors when $D_1$ is odd. These effects are amplified in higher dimensions and larger distances. Interestingly, for a given combination $(p,k)$, increasing the lattice size often results in a larger number of colors. Therefore, it is beneficial if a lattice can be composed with smaller tiles.

There are certain constraints that the tile size must satisfy. First, because the periodicity in the tile must match that of the lattice, a hyper-cubic tile must be used.
Second, the tile needs to be large enough to include an entire $N(x,p,k)$ neighborhood. Otherwise, the neighborhood will wrap-around the boundary and thus require more colors than a larger tile would need.
This means that in dimensions without displacement the length needs to be at least $2k+1$.
The dimension with the displacement should have length at least $2(p+k)+1$. For example, a ($p=8, k=8$)-coloring on a 4D lattice would require a tile of size at least $34 \times 18^3$. 

A third constraint is that the tile dimensions must divide the dimensions of the lattice to ensure a valid coloring. 
In LQCD lattices have dimensions that are a power of two in size, occasionally including a factor of three. Therefore,  the minimum size $34 \times 18^3$ tile of the previous example cannot be used. One solution is to consider tiles with each dimension length being the smallest power of two that is greater than the minimum required length. In the previous example, the tile size required for the ($8, 8$)-coloring on a 4D lattice would be $64 \times 32^3$. 
The drawback of this requirement is that tiles may become too large and some of their dimensions (in particular the one with displacement) may be longer than the size of the actual lattice. In such cases, we may limit the tile size in the offending dimension to $D_i$. 
This ensures a valid coloring, although with possibly a few more colors, but also standardizes the number of tiles we need to pre-compute and store. 
In the example above, if the lattice is of size $32 \times 64^3$, then the size of the ($8, 8$)-coloring tile becomes $32^4$.


\begin{table}[htb]
\centering\tiny
\begin{tabular}{|c||r|r|r|r|r|r|r|r|r|}
\hline
$k$ & \multicolumn{9}{c|}{Displacement} \\ \hline
 & \multicolumn{1}{c|}{0} & \multicolumn{1}{c|}{1} & \multicolumn{1}{c|}{2} & \multicolumn{1}{c|}{3} & \multicolumn{1}{c|}{4} & \multicolumn{1}{c|}{5} & \multicolumn{1}{c|}{6} & \multicolumn{1}{c|}{7} & \multicolumn{1}{c|}{8} \\ \hline \hline
1 & $4^4$ & $8 \times 4^3$ & $8 \times 4^3$ & $16 \times 4^3$ & $16 \times 4^3$ & $16 \times 4^3$ & $16 \times 4^3$ & $32 \times 4^3$ & $32 \times 4^3$ \\ \hline
2 & $8^4$ & $8^4$ & $16 \times 8^3$ & $16 \times 8^3$ & $16 \times 8^3$ & $16 \times 8^3$ & $32 \times 8^3$ & $32 \times 8^3$ & $32 \times 8^3$ \\ \hline
3 & $8^4$ & $16 \times 8^3$ & $16 \times 8^3$ & $16 \times 8^3$ & $16 \times 8^3$ & $32 \times 8^3$ & $32 \times 8^3$ & $32 \times 8^3$ & $32 \times 8^3$ \\ \hline
4 & $16^4$ & $16^4$ & $16^4$ & $16^4$ & $32 \times 16^3$ & $32 \times 16^3$ & $32 \times 16^3$ & $32 \times 16^3$ & $32 \times 16^3$ \\ \hline
5 & $16^4$ & $16^4$ & $16^4$ & $32 \times 16^3$ & $32 \times 16^3$ & $32 \times 16^3$ & $32 \times 16^3$ & $32 \times 16^3$ & $32 \times 16^3$ \\ \hline
6 & $16^4$ & $16^4$ & $32 \times 16^3$ & $32 \times 16^3$ & $32 \times 16^3$ & $32 \times 16^3$ & $32 \times 16^3$ & $32 \times 16^3$ & $32 \times 16^3$ \\ \hline
7 & $16^4$ & $32 \times 16^3$ & $32 \times 16^3$ & $32 \times 16^3$ & $32 \times 16^3$ & $32 \times 16^3$ & $32 \times 16^3$ & $32 \times 16^3$ & $32 \times 16^3$ \\ \hline
8 & $32^4$ & $32^4$ & $32^4$ & $32^4$ & $32^4$ & $32^4$ & $32^4$ & $32^4$ & $32^4$ \\ \hline
9 & $32^4$ & $32^4$ & $32^4$ & $32^4$ & $32^4$ & $32^4$ & $32^4$ & $32^4$ & $32^4$ \\ \hline
10 & $32^4$ & $32^4$ & $32^4$ & $32^4$ & $32^4$ & $32^4$ & $32^4$ & $32^4$ & $32^4$ \\ \hline
\end{tabular}
\caption{Tile sizes for each ($p, k$)-coloring for a $32^3 \times 64$ lattice with the displacement in the first dimension (corresponding to the $z, x, y, t$ dimensions of the application).}
\label{tab:Tiles}
\end{table}

\Cref{tab:Tiles} shows the tiles sizes for different ($p, k$)-colorings chosen with the above policy for a 4-dimensional toroidal lattice of  size $32^3 \times 64$. This is the lattice of our experiments in the next section.
For clarity the table shows the displacement in the first direction, although our LQCD application requires it in the third dimension.

\section{Experiments}
\label{sec:Experiments}

We have implemented our code in C and in MATLAB. The computation of all lattice tiles in \cref{tab:Tiles} was performed with the C code. All tests were run on the Femto subcluster at William \& Mary where each compute node is a 32-core 960 Xeon Skylake with a clock speed of 2.1GHz. The timings for each of the ($p, k$)-colorings on a single thread are shown in \Cref{tab:Times}, but the code can be easily parallelized. While iterating through each node must be sequential in nature to avoid coloring conflicts, gathering the color labels of a single node's neighbors is a read-only process that can be done independently. For example, the maximum number of neighbors each node can have for an $(8, 10)$-coloring is 16,681, allowing for decent speedups. A red-black scheme can also obviously be done in parallel, as the red nodes and black nodes can be separated and colored independently.

\begin{table}[H]
\centering
\tiny
\begin{tabular}{|c||r|r|r|r|r|r|r|r|r|}
\hline
$k$ & \multicolumn{9}{c|}{Displacement} \\ \hline
 & \multicolumn{1}{c|}{0} & \multicolumn{1}{c|}{1} & \multicolumn{1}{c|}{2} & \multicolumn{1}{c|}{3} & \multicolumn{1}{c|}{4} & \multicolumn{1}{c|}{5} & \multicolumn{1}{c|}{6} & \multicolumn{1}{c|}{7} & \multicolumn{1}{c|}{8} \\ \hline \hline
1 & 0.00 & 0.00 & 0.00 & 0.00 & 0.00 & 0.00 & 0.00 & 0.00 & 0.00 \\ \hline
2 & 0.00 & 0.01 & 0.00 & 0.01 & 0.01 & 0.01 & 0.02 & 0.01 & 0.01 \\ \hline
3 & 0.01 & 0.02 & 0.02 & 0.02 & 0.03 & 0.05 & 0.04 & 0.04 & 0.04 \\ \hline
4 & 0.21 & 0.33 & 0.39 & 0.41 & 0.82 & 0.82 & 0.82 & 0.83 & 0.83 \\ \hline
5 & 0.44 & 0.67 & 0.79 & 1.70 & 1.74 & 1.75 & 1.75 & 1.75 & 1.76 \\ \hline
6 & 0.84 & 1.22 & 2.92 & 3.16 & 3.29 & 3.32 & 3.32 & 3.34 & 3.33 \\ \hline
7 & 1.44 & 4.13 & 4.93 & 5.44 & 5.68 & 5.79 & 5.82 & 5.84 & 5.85 \\ \hline
8 & 38.88 & 55.23 & 64.81 & 71.35 & 77.19 & 76.69 & 77.84 & 78.54 & 77.99 \\ \hline
9 & 61.63 & 85.06 & 97.91 & 108.21 & 114.82 & 119.90 & 121.16 & 121.22 & 122.22 \\ \hline
10 & 91.16 & 121.33 & 143.77 & 157.81 & 167.94 & 175.82 & 179.13 & 180.09 & 180.38 \\ \hline
\end{tabular}
\caption{Time (in seconds) to run each ($p, k$)-coloring with tile sizes outlined in \Cref{tab:Tiles} and the resulting number of colors is shown in \Cref{tab:ActualCols}.}
\label{tab:Times}
\end{table}


\subsection{Number of Colors Computed}

As the number of colors equates to the number of linear systems needing to be solved in \Cref{eq:Hutchinson}, we are interested in studying how close the number returned by the greedy algorithm is to the theoretical lower bounds summarized in \cref{eq:pBounds}. As discussed in \Cref{subsec:Tiles}, the lower bounds are for lattices without boundary restrictions so depending on lattice size we expect variability in the deviation from the lower bound.

\begin{table}[htb]
\centering\tiny
\begin{tabular}{|c|r|r|r|r|r|r|r|r|r|}
\hline
$k$ & \multicolumn{9}{c|}{Displacement} \\ \hline
 & \multicolumn{1}{c|}{0} & \multicolumn{1}{c|}{1} & \multicolumn{1}{c|}{2} & \multicolumn{1}{c|}{3} & \multicolumn{1}{c|}{4} & \multicolumn{1}{c|}{5} & \multicolumn{1}{c|}{6} & \multicolumn{1}{c|}{7} & \multicolumn{1}{c|}{8} \\ \hline \hline
1 & 2/2 & 5/3 & 4/4 & 5/3 & 3/3 & 4/3 & 4/4 & 3/3 & 3/3 \\ \hline
2 & 16/9 & 9/6 & 6/5 & 10/6 & 4/4 & 6/4 & 5/3 & 4/3 & 3/3 \\ \hline
3 & 16/16 & 32/23 & 11/10 & 9/7 & 8/8 & 6/5 & 7/4 & 5/4 & 4/4 \\ \hline
4 & 119/41 & 64/40 & 92/37 & 17/14 & 14/9 & 12/10 & 10/6 & 6/5 & 4/4 \\ \hline
5 & 170/66 & 324/91 & 96/64 & 64/51 & 27/18 & 21/11 & 19/12 & 9/7 & 6/6 \\ \hline
6 & 256/129 & 442/142 & 586/141 & 128/88 & 104/65 & 34/22 & 19/13 & 18/14 & 8/8 \\ \hline
7 & 256/192 & 815/255 & 795/218 & 866/192 & 192/112 & 172/79 & 37/26 & 17/15 & 16/16 \\ \hline
8 & 1037/321 & 976/368 & 1024/381 & 1206/294 & 1254/241 & 336/136 & 160/93 & 33/30 & 30/17 \\ \hline
9 & 1298/450 & 2031/579 & 1024/544 & 1760/507 & 1577/370 & 1556/291 & 288/160 & 128/107 & 52/34 \\ \hline
10 & 2220/681 & 2462/790 & 3238/837 & 1922/720 & 2082/633 & 1976/446 & 1954/341 & 256/184 & 264/121 \\ \hline
\end{tabular}
\caption{The first number is the smallest number of colors achieved for distance-$k$, displacement $p$, on the tiles of size as noted in \Cref{tab:Tiles}. The second number is the lower bound for that $(p,k)$ from \cref{eq:pBounds}.}
\label{tab:ActualCols}
\end{table}

\Cref{tab:ActualCols} shows the least amount of colors achieved between natural and red-black orderings for our different $(p, k)$-colorings.
Next to this number is the theoretical lower bound for each $(p, k)$ combination where $p \in \{0, 1,...,8\}$ and $k \in \{1, 2, ..., 10\}$.

\begin{figure}[htb] 
  \centering
  \includegraphics[trim=4.4cm 1.7cm 4cm 2.3cm, clip,width=\linewidth]{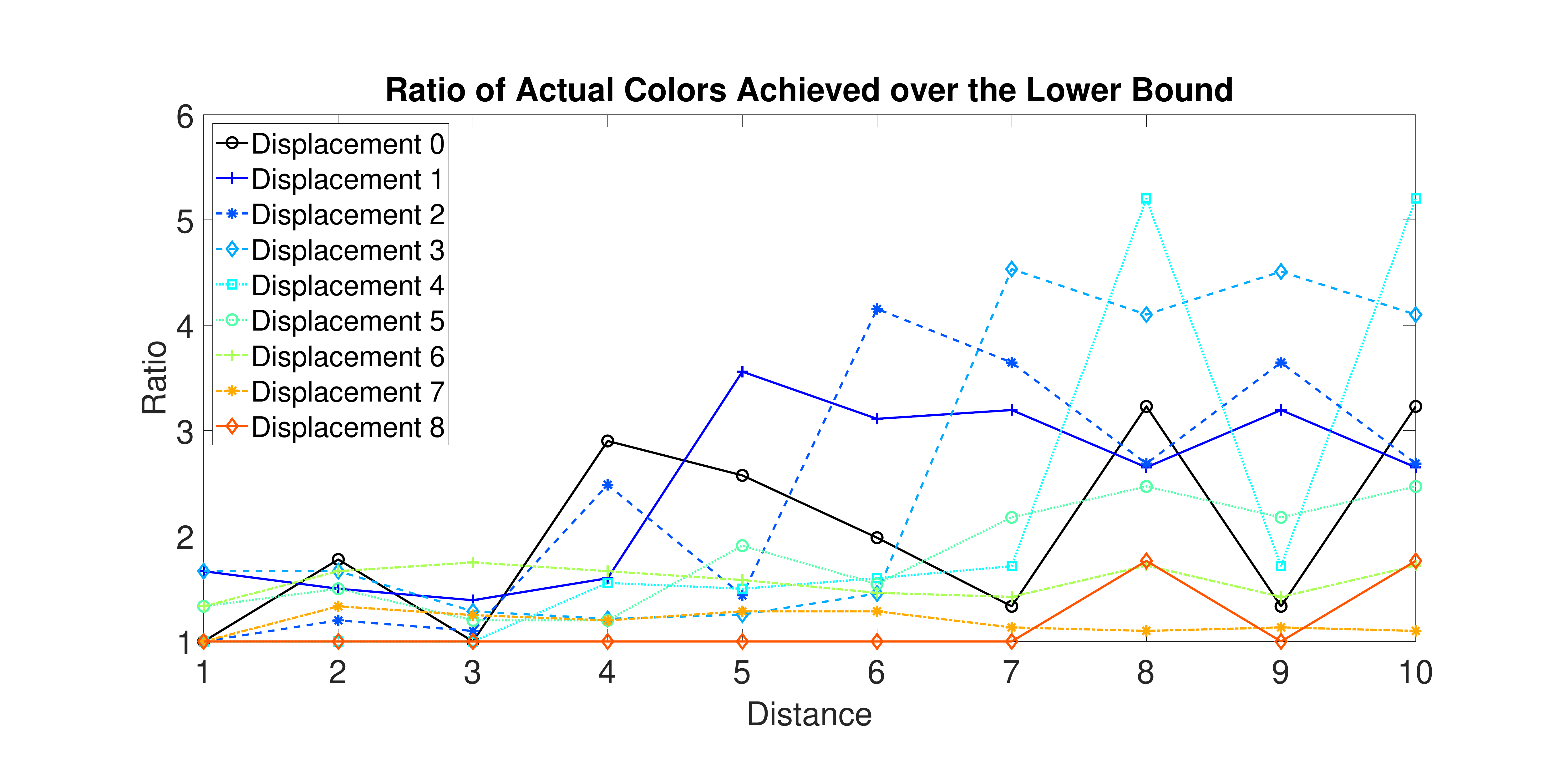} 
  \caption{The ratio of the minimum number of colors achieved with a $(p, k)$-coloring to the theoretical lower bound in \Cref{tab:ActualCols}. Each column is a different displacement.}
  \label{fig:MinColRatio}
\end{figure}

The ratio between the two numbers for all combinations is plotted in \Cref{fig:MinColRatio}.
We observe that when $p \geq k$, the achieved number of colors is very close to the lower bound as the coloring problem becomes one-dimensional, which provides significantly fewer clique constraints. However, once the two displaced neighborhoods begin to overlap, the number of constraints increases and we see the toroidal boundary effects of the tiles. 
In our experiments, $(p,k)$ combinations with ratios less than three and around 200-300 colors seem to work the best.

\subsection{Comparisons to Other Methods}

Based on the tiles outlined in \Cref{tab:Tiles}, we generated probing vectors that were used in trace estimation experiments using the Chroma library from Jefferson Laboratory \cite{Chroma}. The $32^3 \times 64$ lattice generated by Chroma used a Clover fermion action with quark mass of -0.239. The gauge configuration is from the same ensemble listed as Ensemble B in \cite{Deflation}. More details about this ensemble can be found in \cite{ensemble}. 
As suggested in \cite{Deflation}, we deflate with 200 largest singular vectors of $A^{-1}$ which are computed using the PRIMME library \cite{PRIMME}. The solution of each linear system is performed in single precision to relative residual accuracy of 1e-3 with the MG Proto library \footnote{\url{http://jeffersonlab.github.io/qphix} and \url{github.com/jeffersonlab/mg}}, an adaptive multigrid solver, inside of Chroma.

We compare our displacement probing method against the unprobed Hutchinson method and against CP without displacement. 
Because in LQCD each lattice point has 12 degrees of freedom (for all spin-color combinations), all methods perform a probing of these 12 components (called spin-color dilution in the literature \cite{Bali:2009}). This amounts to taking a Kronecker product of each probing or random vector with a $12\times 12$ identity matrix, and thus implies twelve linear systems must be solved for each probing or random vector. We also assume that the matrix $A$ has already been deflated with 200 singular triplets.

Let $v(P_p A^{-1})$ be a shorthand for the variance  \cref{eq:HutchinsonVar} for the matrix $P_p A^{-1}$, where $P_p$ is the permutation matrix that places the elements of $A^{-1}$ corresponding to displacement $p$ in the main diagonal (clearly $P_0 = I$).
The unprobed Hutchinson method is run with $s_1 = 1,000$ Rademacher vectors to estimate the trace and variance of $P_p A^{-1}$. 
For the probing with displacements and the CP methods, let  $H$ be the $N \times m$ matrix with the required $m$ probing vectors as columns, and considering $s_2 = 10 $ Rademacher vectors, construct the $m\times s_2$ vectors $V^{(1)}, \ldots, V^{(s_2)}$  as in \Cref{subsec:DetBias}.
These are used to estimate the trace and variance for each $P_p A^{-1}$.

To compare the methods in a meaningful way we must consider their effect under the same number of linear systems solved. For the Hutchinson method the computed variance of the $s_1$ quadrature values computed in \cref{eq:Hutchinson} provides a good estimation of $v(P_pA^{-1})$. Similarly for the probing variants after $s_2$ Hutchinson steps we expect a good estimation of $v((P_p A^{-1}) \odot HH^T)$. However, each of the $s_2$ stochastic steps of the probing variants solves $m$ linear systems, which implies that the speedup is
\begin{equation}
    \label{eq:Speedup}
    \mbox{Speedup over unprobed Hutchinson} = \frac{v(P_p A^{-1})}{m\times v((P_p A^{-1})\odot HH^T)}.
\end{equation}

\begin{figure}[htb] 
  \centering
  \includegraphics[trim=3.3cm 1.5cm 4.7cm 2.5cm, clip,width=1\linewidth]{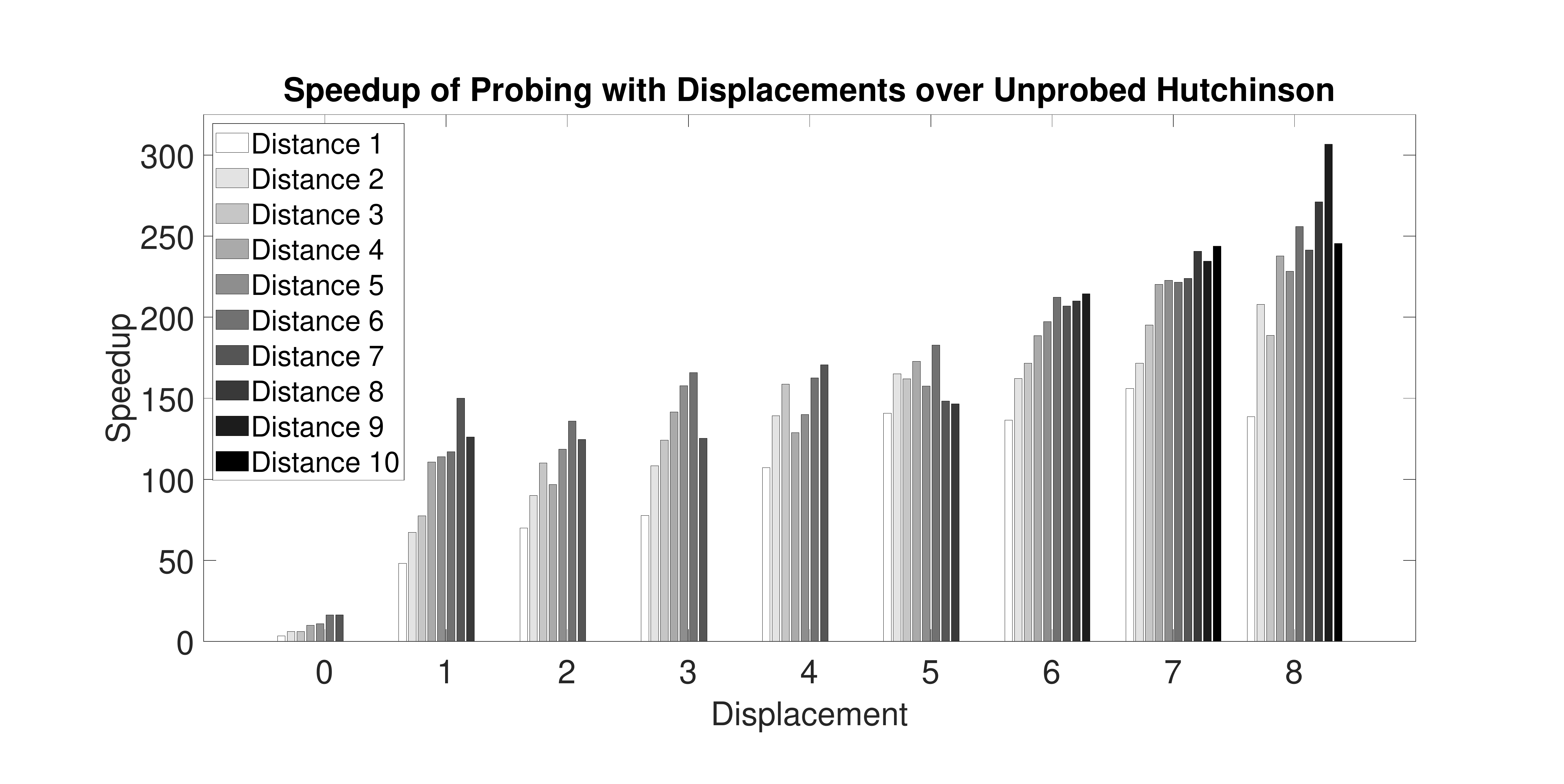} 
  \caption{Speedups of probing with displacements over unprobed Hutchinson using each $(p, k)$-coloring from \Cref{tab:VarTable} to find $tr(P_pA^{-1})$.}
  \label{fig:SpeedupRNvDis}
\end{figure}

\Cref{tab:VarTable} shows the detailed results for trace and variance estimations as well as speedups for Hutchinson and for our new method for different combinations of $p$, $k$.
The speedups for probing with displacements over unprobed Hutchinson are also graphed in \Cref{fig:SpeedupRNvDis}.
We make a few observations.
First, the larger the displacement $p$, the larger the speedup of the new method over unprobed Hutchinson. 
Second, as mentioned before, distance-1 probing has the biggest impact as it removes the main diagonal of $A^{-1}$, as well as the elements at distance-1 away from the main diagonal.
Third, the speedup increases with $k$ but peaks at a certain distance, typically around $k=6$ for smaller displacements and around $k=9$ for larger displacements. 
This is expected as the elements of $A^{-1}$ decay at higher distances making it less beneficial to probe them directly instead of randomly.
Finally, for $p=0$, probing is equivalent to CP and gives a speedup of 16 over unprobed Hutchinson which is slightly better than our previous HP method albeit giving up the hierarchical property.

To solve the problem with displacements, practitioners had previously attempted to use CP or HP  \cite{Alexandrou:2020sml} or a more localized hopping parameter expansion \cite{Zhang:2020dkn}. We want to show the improvements of our method over CP.
Let $m_p$ be the number of probing vectors produced in the ($p,k$)-coloring to form $H_p$.
Clearly the $m_0$ vectors forming $H_0$ are the CP vectors, which could be used to reduce the variance of the estimator for $P_p A^{-1}$.
The speedup of probing with displacements over CP is then,
\begin{equation}
    \mbox{Speedup  over CP} = \frac
    {v((P_p A^{-1}) \odot H_0H_0^T)\times m_0}
    {v((P_p A^{-1}) \odot H_pH_p^T)\times m_p}.
    \label{eq:SpeedupDvND}
\end{equation}

\begin{figure}[htb] 
    \centering
    \includegraphics[trim=3.3cm 1.5cm 4.7cm 2.5cm, clip,width=1\linewidth]{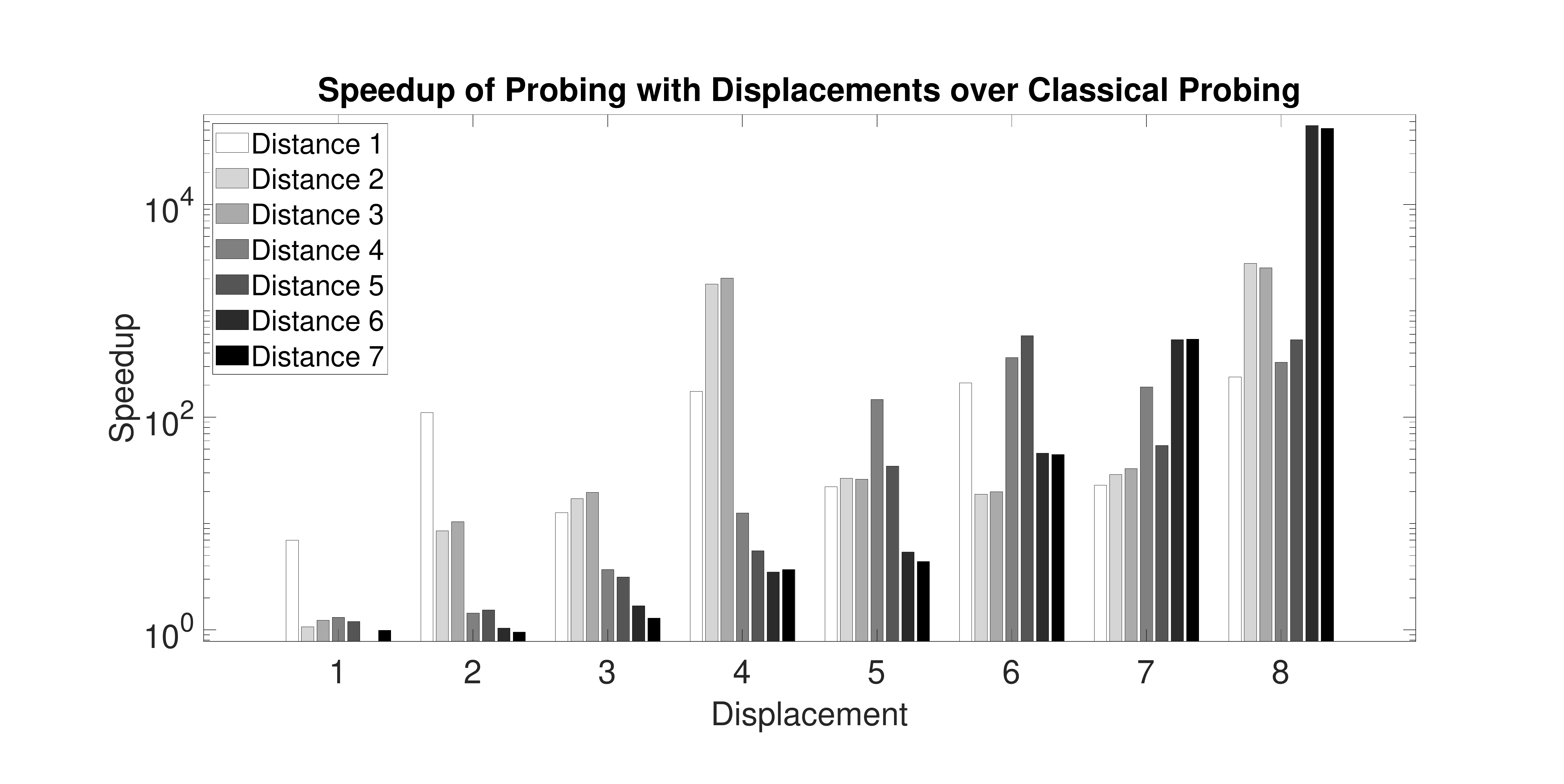} 
  \caption{Speedup of probing with displacements with $(p,k)$-colorings over classical probing with $(0, k)$-colorings to find $tr(P_pA^{-1})$ using \cref{eq:SpeedupDvND}.}
  \label{fig:DisvCP}
\end{figure}

In \Cref{fig:DisvCP} we can see this speedup increasing with displacement, although for small displacements it decreases with distance. This is because CP builds its neighborhood outward from the new diagonal, so it can only eliminate the original main diagonal when $k\geq p$. Even then, as the displacement grows the number of colors the new method needs to achieve a distance-$k$ coloring becomes much smaller. For example, a $(0, 7)$-coloring uses 256 colors, while an $(8, 7)$-coloring only uses 16. Therefore, even if CP does remove the high-magnitude elements eventually, it can take many more probing vectors to do so.

\subsection{Using one coloring for all displacements}

\Cref{thm:Clearance} showed that a $(p_0, k_0)$-coloring would clear all nodes up to distance $k = max(0, k_0-|p_0-p|)$ for a displacement of $p$. \Cref{tab:Clearance} confirms this experimentally for the $(8, 10)$-coloring but also shows how many nodes are {\em not} annihilated beyond the distance described by the theorem. To obtain this, for each pair of $(p,k),$ $ p=0,\ldots, 8, k=1,\ldots, 12$, we go through every node $x$ in the lattice and compute the percentage of nodes exactly at distance-$k$ from $x^+$ or $x^-$ that share the same color label as $x$. These are distance-$k$ neighbors that are not annihilated by the $(8,10)$-coloring. We report the average of this percentage over all $N$ nodes. When the percentage is $0.00$, it means that distance is ``cleared", i.e., all nodes of that distance are annihilated from the variance.

\begin{table}[htb]
\centering\footnotesize
\begin{tabular}{|c||r|r|r|r|r|r|r|r|r|}
\hline
$k$ & \multicolumn{9}{c|}{Displacement} \\ \hline 
 & \multicolumn{1}{c|}{0} & \multicolumn{1}{c|}{1} & \multicolumn{1}{c|}{2} & \multicolumn{1}{c|}{3} & \multicolumn{1}{c|}{4} & \multicolumn{1}{c|}{5} & \multicolumn{1}{c|}{6} & \multicolumn{1}{c|}{7} & \multicolumn{1}{c|}{8} \\ \hline \hline
1 & 0.00 & 0.00 & 0.00 & 0.00 & 0.00 & 0.00 & 0.00 & 0.00 & 0.00 \\ \hline
2 & 0.00 & 0.00 & 0.00 & 0.00 & 0.00 & 0.00 & 0.00 & 0.00 & 0.00 \\ \hline
3 & 4.55 & 0.00 & 0.00 & 0.00 & 0.00 & 0.00 & 0.00 & 0.00 & 0.00 \\ \hline
4 & 9.38 & 1.35 & 0.00 & 0.00 & 0.00 & 0.00 & 0.00 & 0.00 & 0.00 \\ \hline
5 & 3.33 & 3.41 & 0.63 & 0.00 & 0.00 & 0.00 & 0.00 & 0.00 & 0.00 \\ \hline
6 & 0.00 & 1.40 & 1.76 & 0.35 & 0.00 & 0.00 & 0.00 & 0.00 & 0.00 \\ \hline
7 & 2.52 & 0.00 & 0.78 & 1.05 & 0.22 & 0.00 & 0.00 & 0.00 & 0.00 \\ \hline
8 & 4.69 & 1.29 & 0.00 & 0.49 & 0.69 & 0.15 & 0.00 & 0.00 & 0.00 \\ \hline
9 & 2.01 & 2.56 & 0.79 & 0.00 & 0.33 & 0.47 & 0.10 & 0.00 & 0.00 \\ \hline
10 & 0.00 & 1.16 & 1.64 & 0.53 & 0.00 & 0.24 & 0.34 & 0.07 & 0.00 \\ \hline
11 & 1.66 & 0.00 & 0.77 & 1.14 & 0.38 & 0.00 & 0.18 & 0.26 & 0.06 \\ \hline
12 & 3.13 & 1.05 & 0.00 & 0.54 & 0.83 & 0.29 & 0.00 & 0.13 & 0.20 \\ \hline
\end{tabular}
\caption{The average percentage of neighbors at exactly distance-$k$ that do not get eliminated from the trace estimator when using a ($8, 10$)-coloring to find other displacements. The lattice size used is $32^3 \times 64$ with a tile size of $32^4$.}
\label{tab:Clearance}
\end{table}

The presence of zeros for any $k \leq 10-|p-8|$ confirms \Cref{thm:Clearance}. For each $p$, we also observe a zero at  distances $4i + (k_0-|p_0-p|), \forall i \in \mathbb{Z}_+$ which may be attributed to wrap-around effects and/or the red-black ordering that was used for the $(8, 10)$-coloring. 
More importantly, however, the percentages of uncleared elements at larger distances is still very small, often less than 1\%. This is because a coloring annihilates the distance-$k$ neighbors of all nodes of the same color. For example, if $x_1$ and $x_2$ have the same color, some of the neighbors of $x_1$ may be longer distance neighbors of $x_2$ but they are annihilated for this $k$.
 
\begin{figure}[htb]
\centering
    \includegraphics[trim=3.23cm 3.3cm 3.5cm 4.5cm, clip,width=\linewidth]{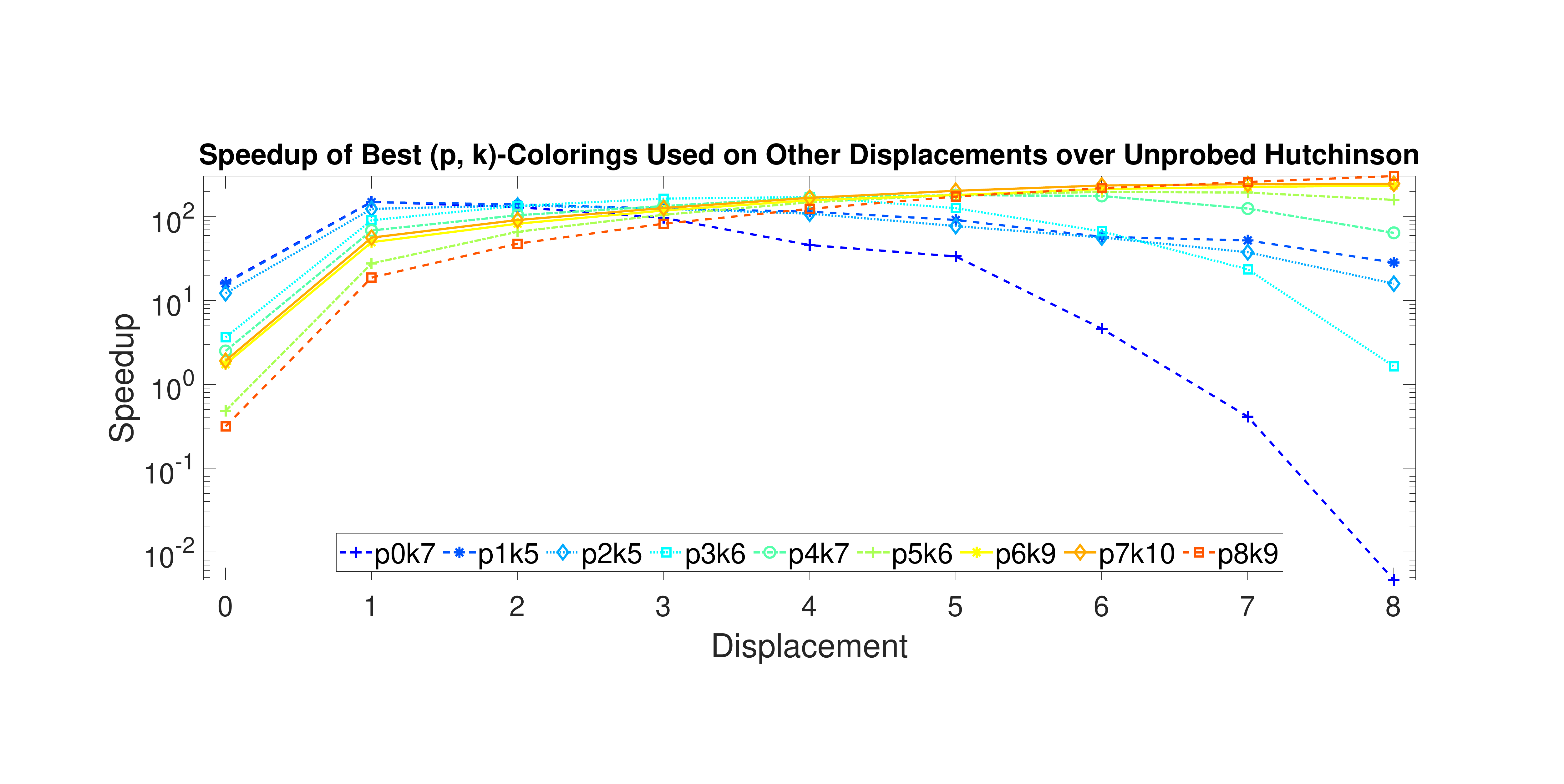} 
   \caption{The speedups over unprobed Hutchinson for each $(p, k_p)$-coloring to find $tr(P_nA^{-1})$, $\forall n \in \{0, ..., 8\}$} 
   \label{fig1:SpeedupBestColsOtherDis} 
\end{figure}

Next, we study the effects of this strategy on variance reduction. For each $p$, we take the $(p, k_p)$-coloring that gives the best speedup over unprobed Hutchinson (from \Cref{fig:SpeedupRNvDis}) and use it to find the variance $v((P_{n}A^{-1})\odot H_pH_p^T)$  for all other displacements $n=0,\ldots, 8$. \Cref{fig1:SpeedupBestColsOtherDis} shows nine lines, one for each $p$, plotting its speedup over the Hutchinson method for all $n$.
Each line achieves its maximum speedup at $n=p$ or for smaller $p$, at $n=p+1$ . It is unclear why this happens for smaller $p$, e.g.,  most pronounced for the $(0, 7)$-coloring, but it may have to do with the symmetrization. More importantly, the speedup does not reduce as steeply away from $p$ as \Cref{thm:Clearance} would suggest because these colorings work very well for nearby displacements and still work well for more distant ones as described in \Cref{tab:Clearance}.

The above results help ascertain the efficiency of the approach, but they cannot help determine which coloring should be used to perform all displacement experiments. There are two reasons. First, the speedups reported depend on the number of probing vectors used. For example, the $(8, 9)$-coloring obtains a speedup of $300$ at $p=8$ but it's because it uses only 52 colors. Its variance is actually four times larger than that of $(7,10)$-coloring which however uses 250 vectors and thus gets a  lower speedup of $250$. For a more accurate answer, the $(7,10)$-coloring would be a better choice. 

Second, a smaller variance is only meaningful relative to the value of the trace, and traces for different displacements vary significantly. In \Cref{tab:VarTable} we see that a variance of 3.275 for the $(0,4)$-coloring gives 5 digits of accuracy for the trace of $p=0$, while a variance of 2.332 for the $(8,9)$-coloring hardly attains a digit for the trace of $p=8$.

\begin{figure}[htb]
\centering
    \includegraphics[trim=3.23cm 3.3cm 3.5cm 4.4cm, clip,width=\linewidth]{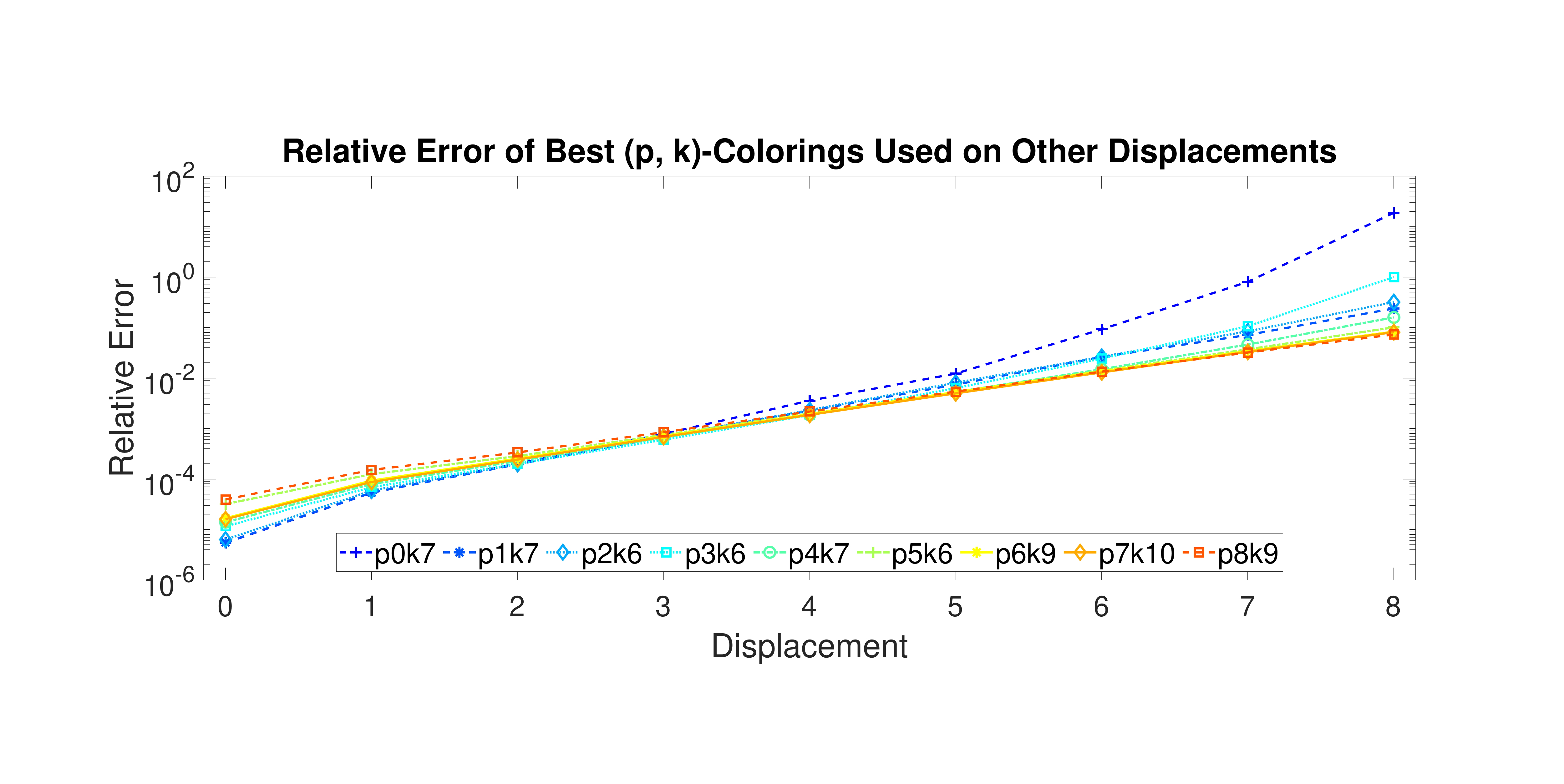} 
    \caption{The relative error \cref{eq:NormRE} for each $(p, k_p)$-coloring used to find $tr(P_nA^{-1})$, $\forall n \in \{0, ..., 8\}$.}
    \label{fig2:REBestCols} 
\end{figure}

Therefore, to compare colorings over different displacements we introduce the normalized relative error metric which normalizes with respect to both the trace and the number of probing vectors needed.
As before, for each $p$ we pick the $(p, k_p)$-coloring with the best speedup over unprobed Hutchinson. Let $m_p$ be the number of colors it requires, and let $M$ be the maximum number of colors over all colorings being compared (in this case, $M = 815$).
Then, for all $n=0,\ldots, 9$,  the normalized relative error is given by,
\begin{equation}
    \label{eq:NormRE}
    \frac{\sqrt{v((P_{n}A^{-1})\odot H_pH_p^T) \frac{m_p}{M}} 
         }{tr(P_nA^{-1})}.
\end{equation}
The normalization to $M$ ensures all colorings are compared as if they use the same number of probing vectors. The results of this shown in \Cref{fig2:REBestCols}.

The fact that the trace decreases significantly in higher displacements provides a much clearer evaluation picture. For $2\leq n \leq 6$, all $(p,k_p)$-colorings have similar normalized relative errors. However, the colorings from larger displacements, e.g., $(7,10)$ or $(8,9)$, yield at least 1 to 2.5 digits better accuracy for the same amount of work than colorings from small displacements. Because for displacements less than 4 the errors are already very small, the effort must be focused on the small traces of higher displacements. Therefore, it is best to use the $(7,10)$-coloring for all displacements, and increase its distance if needed.

\section{Conclusion}
\label{sec:Conclusion}

We have extended the idea of probing for variance reduction of the Hutchinson's trace estimator to the case of permuted matrices and in particular when this permutation corresponds to a lattice displacement $p$. This has an important application on disconnected diagrams in LQCD.
The method works by computing a distance-$k$ coloring not of the original neighborhood of each lattice point $x$ but rather the points within a distance $k$ around centers  $x \pm p$.

We have provided a lower bound of the number of colors needed for a particular $(p, k)$-coloring, and discussed the impact of the lattice size on the number of colors achieved. We have also studied theoretically and experimentally the effect of using a single $(p,k)$-coloring for displacements other than $p$. We have shown that the variance reduction of using probing with displacements is orders of magnitude lower than solely using random noise vectors or than using classical probing that does not take the displacement into consideration. Also, as expected, the trace is smaller as the displacement increases which means that a $(p,k)$-coloring for larger $p$ needs to be computed and then reused for lower $p$. This practically gives an additional 10-fold speedup for the LQCD application.

A few open problems could be considered further. The greedy orderings we considered in the greedy coloring approach did not vary substantially in the resulting number of colors, staying within a factor of 3 from the lower bound. It is unclear whether a different ordering can provide considerable reduction in the current number of colors. 
A second direction is to study  the effect of the lattice or tile size to the coloring. Understanding this theoretically rather than experimentally, and providing also a lower bound on the number of colors based on a finite lattice size could be useful in understanding the limitations of the current approach. Finally, it is worth extending the theory and algorithms to the case where the decay of the elements in the matrix inverse depends on the L2 distance, which is closer to what LQCD theory predicts for long range distances.

\bibliographystyle{plain}
\section*{Appendix A} 
\mbox{}\\

{\em Proof for \cref{thm:1}:}
 From the definition of $x^+, x^-$, the assumption $p\geq k$ implies $| x_1 - x^{+}_1|=| x_1 - x^{-}_1| \geq k$, 
and  $\sum_{i=2}^d | x_i-x_i^+|=
      \sum_{i=2}^d | x_i-x_i^-| =0$.
Let $y\in {Z}^d_{\infty}$ with $y\not = x$ and $y_1 = x_1$. Since  $| y_1 - x_1^+ | \geq k$,  $\|y-x^+\|_1  = | x_1 - x_1^+| + 
\sum_{i=2}^d | x_i-x_i^+| > k$. The same argument applies for the distance from $x^{-}$. Therefore, $y \notin N(x, p, k)$.
\hfill $\square$

{\em Proof for \cref{thm:2}:}
From \Cref{thm:1}, if $p = k$, the coloring problem reduces to coloring in one dimension.
Then the neighborhood definition covers $4k+1$ indices in the first dimension, from $-2k$ to $+2k$. The maximal clique size of the $2k$-distance graph of these nodes is the 1D unit ball of half the size which includes indices from $-k$ to $+k$. The size of this clique is exactly $2k+1$ nodes which is also the number of colors needed. 
\hfill $\square$

{\em Proof for \cref{thm:3}:}
As previously noted, if $p \geq k$, the coloring problem gets reduced to one dimension. Consider two nodes on $\mathbb{Z}^1_{\infty}$ that are exactly $2p$ links apart, i.e., node $i$ and $i+2p$. Because $p>k$, node $i+2p$ does not belong in $N(i,p,k)$ and thus the two nodes can take the same color. Therefore, the problem reduces to coloring a tile of $2p$ consecutive nodes that can then repeat to color the entire $\mathbb{Z}^1_{\infty}$ lattice.

Consider $2p$ consecutive nodes, 0 to $2p-1$.
The first $p-k$ can share color 1 as they do not belong in each other's neighborhood.
The following $p-k+1,\ldots ,2(p-k)$ nodes have at least one of the nodes in the first group as neighbor so they must take a different color, say 2. Similarly, every $p-k$ group of nodes must take a different color. The last node has neighbors $i\geq 2p-p-k = p-k$, so it cannot reuse any color including color 1 because the tile needs to repeat.
Then the total number of colors is the partitioning of $2p$ nodes in $p-k$ groups.
Note that $\lceil \frac{2p}{p-k} \rceil = 
\lceil \frac{2(k+p-k)}{p-k} \rceil = 
\lceil \frac{2k}{(p-k)}\rceil +2$.
\hfill $\square$

{\em Proof for \cref{thm:evenC}:}
First note that $\alpha - \beta = p$.
Let $x, y \in C(d,\alpha,\beta)$. Then the following hold: 
\begin{align}
    |x_1|+\sum_{i=2}^d |x_i| \leq \alpha, &\qquad
    |y_1| + \sum_{i=2}^d |y_i| \leq \alpha
    \label{eq:condition1} \\
   \sum_{i=2}^d |x_i| \leq \beta, &\qquad
   \sum_{i=2}^d |y_i| \leq \beta.
  \label{eq:condition2}
\end{align}

WLOG assume $x_1 \leq y_1$.
Then it is sufficient to show that $x$ belongs in the left neighborhood around $y^-$, i.e.,
$x \in N(y^-, 0, k)$ or 
$    | y^-_1 - x_1| + \sum_{i=2}^n | y_i - x_i| \leq k
$.

We distinguish two cases for the distance between $x_1$ and $y_1$.
\begin{itemize}
    \item[(a)] $y_1 - x_1 \geq p > 0$. Then using \cref{eq:condition1}, we have 
$ | y^-_1 - x_1| + \sum_{i=2}^n | y_i - x_i| =  y^-_1 - x_1 + \sum_{i=2}^n | y_i - x_i| \leq 
| y_1 | + | x_1| - p + \sum_{i=2}^n | y_i| + \sum_{i=2}^n | x_i| \leq k+p-p = k$.

\item[(b)] $0 \leq y_1 - x_1 < p$. Then using \cref{eq:condition2}, we have: $| y^-_1 - x_1| + \sum_{i=2}^n | y_i - x_i| =  -y_1+x_1 + p  + \sum_{i=2}^n | y_i - x_i| 
\leq p+ \sum_{i=2}^n | y_i| + \sum_{i=2}^n | x_i| \leq p + k -p=k$.
\end{itemize}
\hfill $\square$

{\em Proof for \cref{thm:oddC}:}
For brevity we denote $C=C(d,\alpha,\beta)$. Let $x, y \in C'$, i.e., they belong in one of the sets $C,T,S$.
Because of symmetry, we consider the following pairs of conditions for $(x,y)$:
$(C,C)$, $(T,T)$, $(S,S)$, 
$(C,T)$, $(C,S)$, $(T,S)$.

Notice that the set $C$ is the clique obtained by $k'=k-1$ and $p$. Then, case ($C,C)$ is covered by \Cref{thm:evenC} which bounds the (displaced) distance of any two points in $C$ by $k'=k-1 < k$. 
This observation can be used to show similarly the cases $(C,T)$ and $(C,S)$. 
Specifically for $(C,T)$, $x\in C$ and any $y\in T$ will be exactly at distance 1 from some point in $C$, which means  $\|x-y\| \leq k' + 1 = k$.
For $(C,S)$, a $y\in S$ is also at distance 1 from any point in $C$ by extending the first dimension.

\noindent
As in \Cref{thm:evenC}, we assume $x_1 \leq y_1$ and show that $x$ belongs in the left neighborhood around $y^-$, i.e.,
$x \in N(y^-, 0, k)$ or
$   \delta = | y^-_1 - x_1| + \sum_{i=2}^d | y_i - x_i| \leq k. 
$
We also use the following property of absolute values,
\begin{align}
 |f-g| - |f| -|g| = 
 \left\{
 \begin{array} {cl}
 0,  & \mbox{if } fg \leq 0,\\
 -2\min(|f|,|g|), & \mbox{otherwise}.
 \end{array}
 \right. 
 \label{eq:abs}
\end{align}

\noindent
$\bullet$ {\em Case $(T,T)$:} \\
Using the last two conditions of \cref{cond:2}, the corresponding part of \cref{eq:abs} for $x_2, y_2$, and
$2\beta = k-p-1$ we have,
\begin{align*}
\sum\nolimits_{i=2}^d |y_i - x_i| 
& \leq |y_2 -x_2| + \sum\nolimits_{i=3}^d |y_i|  + \sum\nolimits_{i=3}^d |x_i|   \\
& \leq |y_2 -x_2| + (\beta +1 - |y_2|) +
    (\beta + 1 -|x_2|) \\
& = 2\beta + 2 -2\min(|x_2|,|y_2|) \leq k-p-1.
\end{align*}
Then $\delta  = |y_1^- - x_1| + \sum_{i=2}^d |y_i - x_i|  \leq |y_1 - p - x_1| + k-p-1.$
Using the first condition in \cref{cond:2} we have,\\
If $y_1-p \geq x_1$ then $\delta \leq y_1 - p - x_1 + k-p - 1
\leq (p) -p + (p -1) + (k-p-1)  = k - 2 < k.$\\
If $y_1-p < x_1$ or $y_1-x_1< p$ then $\delta \leq x_1 -y_1 + p + (k-p-1) \leq p + k-p-1 < k.$

\noindent
$\bullet$ {\em Case $(S,S)$:}\\
Again we prove $x \in N(y^-, 0, k)$. 
Based on the conditions in \cref{cond:3},
$\sum_{i=3}^d |x_i| = \alpha + 1 - x_1 - |x_2|$, and
$\sum_{i=3}^d |y_i| = \alpha + 1 - y_1 - |y_2|$, and since $2\alpha = k+p-1$ we have, 
\begin{align*}
    \delta
& \leq |y_1^- - x_1| + |y_2 - x_2| + \sum\nolimits_{i=3}^d |y_i| + \sum\nolimits_{i=3}^d |x_i|\\
& = |y_1^- - x_1| + |y_2 - x_2| + 2 \alpha + 2 - 
|x_2|-|y_2| - x_1 -y_1\\
& =  k+p+1 + (|y_1 - p - x_1|- x_1 -y_1 ) + (|y_2 - x_2| -|x_2|-|y_2|) \\
& \leq k+p+1 + (p + |y_1 - x_1|- x_1 -y_1 ) -2\min(|x_2|,|y_2|)\\
& \leq k+2p+1 -2 \min(|x_1|,|y_1|)
    -2\min(|x_2|,|y_2|)\\ 
&   = k + 2p +1 -2(p+1) = k-1 < k.
\end{align*}
    
\noindent
$\bullet$ {\em Case $(T,S)$:}\\  
Let $x\in T$, $y\in S$. From the defining conditions, $x_1\leq p < y_1$. We work similarly with the previous cases, replacing the $\sum_{i=3}^d$, and noting that $\alpha + \beta = k-1$,
\begin{align*}
    \delta
& \leq |y_1^- - x_1| + |y_2 - x_2| +
    (\alpha + 1-|y_1| - |y_2|)
  + (\beta + 1 - |x_2|)\\
& = k+1  + 
    (|y_1 -p - x_1| -|y_1|) +
    (|y_2 - x_2| -|y_2|-|x_2|)\\
& \leq  k+1  + 
    (|y_1 -p - x_1| -|y_1|) .
\end{align*}
If $y_1 -p \geq x_1 $, then
$|y_1 -p - x_1| -|y_1|
 = y_1 -p - x_1 -y_1 = 
  -p - x_1 \leq -p +(p-1) = -1$. Thus
  $\delta \leq k.$\\
If $y_1 -p \leq x_1 $, then
$|y_1 -p - x_1| -|y_1|
 = x_1 + p - y_1 - y_1 
 \leq 2p - 2(p+1) = -2.$
 Thus, $\delta \leq k-1 < k$.
\hfill $\square$

{\em Proof for \cref{thm:Clearance}:}
We consider only the $p+\lambda$ case as the $p-\lambda$ has a similar proof. Because of symmetry, we also consider only the positive displacements $x^+$ and $y^+$ from \cref{eq:x+x-}.
It is sufficient to show that if 
$x\in N(\mathbf{0}, p+\lambda, k-\lambda)$, then $x \in N(y^+,0,k)$. From \cref{eq:Neighborhood} we have
$    \sum_{i=2}^n \left| x_i\right| + \left| x_1 - (p+\lambda)\right| \leq k-\lambda
$.
We distinguish the following cases. 
\begin{itemize}
    \item[(a)] If $x_1-p \geq \lambda$, then also $x_1 \geq p$, and thus
$ \sum_{i=2}^n \left| x_i\right| + x_1 - p- \lambda \leq k - \lambda 
\Rightarrow \sum_{i=2}^n \left| x_i\right| + |x_1 -p| \leq k  \Rightarrow x \in N(\mathbf{0}, p, k)$.
    \item[(b)] If $x_1 <  p + \lambda$, then  
$ \sum_{i=2}^n \left| x_i \right| - x_1 + p \leq k - 2\lambda$. 
We distinguish two sub-cases. 
    \begin{itemize}
        \item[(b.1)] If $x_1\leq p$, then 
$ \sum_{i=2}^n \left| x_i \right| + | x_1 - p|  \leq k - 2\lambda \leq k
\Rightarrow x \in N(\mathbf{0}, p, k)$.
        \item[(b.2)] If $x_1 > p$ and since $x_1-p < \lambda$, 
then $ \sum_{i=2}^n \left| x_i \right| +  p-x_1  \leq k - 2\lambda  \Rightarrow
\sum_{i=2}^n \left|  x_i \right| + x_1 - p \leq k-2\lambda +2(x_1-p) < k-2\lambda + 2\lambda = k
\Rightarrow x \in N(\mathbf{0}, p, k)$.
    \end{itemize}
\end{itemize}
\hfill $\square$

{\bf Appendix B}\\
\begin{table}[htb]
\centering
\scriptsize
\begin{tabular}{|c|c|r||r|r|r|r|}
\hline
\multicolumn{1}{|r|}{} & \multicolumn{1}{r|}{} & \multicolumn{1}{c||}{} & \multicolumn{1}{c||}{1,000 RNVs w/o Probing} & \multicolumn{2}{c||}{10 RNVs w/ Probing} &  \\ \hline
$p$ & $k$ & \multicolumn{1}{c||}{Approx. Trace} & \multicolumn{1}{c||}{Variance} & \multicolumn{1}{c|}{Colors} & \multicolumn{1}{c||}{Variance} & \multicolumn{1}{c|}{Speedup} \\ \hline \hline
\multirow{7}{*}{0} & 1 & \multirow{7}{*}{6,339,643.7} & \multirow{7}{*}{249,827.7} & 2 & 35,075.3 & 3.56 \\
 & 2 &  &  & 16 & 2,502.5 & 6.24 \\
 & 3 &  &  & 16 & 2,501.2 & 6.24 \\
 & 4 &  &  & 119 & 209.6 & 10.02 \\
 & 5 &  &  & 170 & 134.2 & 10.95 \\
 & 6 &  &  & 256 & 59.4 & 16.43 \\
 & 7 &  &  & 256 & 59.1 & 16.50 \\ \hline
\multirow{8}{*}{1} & 1 & \multirow{8}{*}{652,636.1} & \multirow{8}{*}{2,341,455.9} & 5 & 9,721.2 & 48.17 \\
 & 2 &  &  & 9 & 3,861.4 & 67.38 \\
 & 3 &  &  & 32 & 943.5 & 77.55 \\
 & 4 &  &  & 64 & 330.2 & 110.78 \\
 & 5 &  &  & 324 & 63.4 & 113.94 \\
 & 6 &  &  & 442 & 45.2 & 117.08 \\
 & 7 &  &  & 815 & 19.1 & 150.13 \\
 & 8 &  &  & 976 & 19.0 & 126.21 \\ \hline
\multirow{7}{*}{2} & 1 & \multirow{7}{*}{185,764.9} & \multirow{7}{*}{2,360,726.0} & 4 & 8,415.7 & 70.13 \\
 & 2 &  &  & 6 & 4,362.0 & 90.20 \\
 & 3 &  &  & 11 & 1,949.6 & 110.08 \\
 & 4 &  &  & 92 & 264.9 & 96.87 \\
 & 5 &  &  & 96 & 207.2 & 118.70 \\
 & 6 &  &  & 586 & 29.6 & 135.95 \\
 & 7 &  &  & 795 & 23.8 & 124.73 \\ \hline
\multirow{7}{*}{3} & 1 & \multirow{7}{*}{56,047.8} & \multirow{7}{*}{2,364,612.0} & 5 & 6,076.0 & 77.84 \\
 & 2 &  &  & 10 & 2,180.4 & 108.45 \\
 & 3 &  &  & 9 & 2,115.0 & 124.22 \\
 & 4 &  &  & 17 & 982.1 & 141.63 \\
 & 5 &  &  & 64 & 234.1 & 157.82 \\
 & 6 &  &  & 128 & 111.4 & 165.89 \\
 & 7 &  &  & 866 & 21.8 & 125.41 \\ \hline
\multirow{7}{*}{4} & 1 & \multirow{7}{*}{17,893.6} & \multirow{7}{*}{2,388,516.5} & 3 & 7,420.1 & 107.30 \\
 & 2 &  &  & 4 & 4,285.6 & 139.33 \\
 & 3 &  &  & 8 & 1,880.1 & 158.81 \\
 & 4 &  &  & 14 & 1,323.9 & 128.87 \\
 & 5 &  &  & 27 & 631.8 & 140.02 \\
 & 6 &  &  & 104 & 141.2 & 162.67 \\
 & 7 &  &  & 192 & 72.9 & 170.66 \\ \hline
\multirow{8}{*}{5} & 1 & \multirow{8}{*}{6,059.5} & \multirow{8}{*}{2,358,840.1} & 4 & 4,186.9 & 140.85 \\
 & 2 &  &  & 6 & 2,379.5 & 165.22 \\
 & 3 &  &  & 6 & 2,425.6 & 162.08 \\
 & 4 &  &  & 12 & 1,137.2 & 172.86 \\
 & 5 &  &  & 21 & 712.8 & 157.58 \\
 & 6 &  &  & 34 & 379.3 & 182.90 \\
 & 7 &  &  & 172 & 92.4 & 148.40 \\
 & 8 &  &  & 332 & 48.4 & 146.65 \\ \hline
\multirow{9}{*}{6} & 1 & \multirow{9}{*}{2,183.3} & \multirow{9}{*}{2,392,640.1} & 4 & 4,375.9 & 136.70 \\
 & 2 &  &  & 5 & 2,948.7 & 162.29 \\
 & 3 &  &  & 7 & 1,990.8 & 171.69 \\
 & 4 &  &  & 10 & 1,267.6 & 188.75 \\
 & 5 &  &  & 19 & 638.1 & 197.36 \\
 & 6 &  &  & 19 & 592.9 & 212.40 \\
 & 7 &  &  & 37 & 312.5 & 206.92 \\
 & 8 &  &  & 160 & 71.2 & 210.12 \\
 & 9 &  &  & 288 & 38.7 & 214.56 \\ \hline
\multirow{10}{*}{7} & 1 & \multirow{10}{*}{836.9} & \multirow{10}{*}{2,378,138.9} & 3 & 5,081.5 & 156.00 \\
 & 2 &  &  & 4 & 3,463.9 & 171.64 \\
 & 3 &  &  & 5 & 2,435.2 & 195.31 \\
 & 4 &  &  & 6 & 1,798.8 & 220.35 \\
 & 5 &  &  & 9 & 1,185.9 & 222.82 \\
 & 6 &  &  & 18 & 596.2 & 221.62 \\
 & 7 &  &  & 17 & 624.3 & 224.07 \\
 & 8 &  &  & 33 & 299.3 & 240.76 \\
 & 9 &  &  & 128 & 79.2 & 234.68 \\
 & 10 &  &  & 256 & 38.1 & 243.95 \\ \hline
\multirow{10}{*}{8} & 1 & \multirow{10}{*}{339.3} & \multirow{10}{*}{2,381,007.2} & 3 & 5,719.0 & 138.78 \\
 & 2 &  &  & 3 & 3,814.9 & 208.04 \\
 & 3 &  &  & 4 & 3,151.8 & 188.86 \\
 & 4 &  &  & 4 & 2,502.7 & 237.85 \\
 & 5 &  &  & 6 & 1,737.5 & 228.40 \\
 & 6 &  &  & 8 & 1,162.5 & 256.02 \\
 & 7 &  &  & 16 & 616.1 & 241.53 \\
 & 8 &  &  & 30 & 292.7 & 271.18 \\
 & 9 &  &  & 52 & 149.2 & 306.80 \\
 & 10 &  &  & 264 & 36.7 & 245.51 \\ \hline
\end{tabular}
\caption{The estimation of traces and variances for 1,000 RNVs run without probing for different values of $p$ and $k$ compared to  probing with displacements and 10 RNVs.}
\label{tab:VarTable}
\end{table}

\end{document}